\documentclass[11pt]{article}
 \pdfoutput=1
\usepackage{amsfonts}
\usepackage{graphicx}
\usepackage{amsmath,bm}
\usepackage{amssymb}
\usepackage{latexsym}
\usepackage{mathrsfs}
\usepackage{color}
\usepackage{slashed}
\definecolor{red}{rgb}{1,0,0}
\usepackage{cancel}
\usepackage[nosort]{cite}
\usepackage{color}
\usepackage{booktabs}
\usepackage{multirow}
\usepackage[font=small]{caption}
\usepackage{placeins} %

\usepackage{soul} %

\usepackage{setspace} %

\definecolor{red}{rgb}{1,0,0}

\makeatletter
\def\section{\@startsection {section}{1}{\z@}{-3.5ex plus -1ex minus
 -.2ex}{2.3ex plus .2ex}{\large\bf}}
\def\subsection{\@startsection{subsection}{2}{\z@}{-3.25ex plus -1ex
minus -.2ex}{1.5ex plus .2ex}{\normalsize\bf}}
\makeatother
\makeatletter

\@addtoreset{equation}{section}

\makeatother

\def\bel{\begin{equation}\begin{aligned}}
\def\eel{\end{aligned}\end{equation}}

\def\bea{\begin{eqnarray}} \def\eea{\end{eqnarray}}
\def\be{\begin{equation}} \def\ee{\end{equation}} 
\def\nn{\nonumber}

\newcommand\gt{\widetilde g}
\newcommand\mt{{\widetilde m}}
\newcommand\Mt{{\widetilde M}}

\newcommand\Lt{{\widetilde \Lambda}}

\newcommand\Lagt{\widetilde{\cal L}}

\newcommand{\promille}{%
  \relax\ifmmode\promillezeichen
        \else\leavevmode\(\mathsurround=0pt\promillezeichen\)\fi}
\newcommand{\promillezeichen}{%
  \kern-.05em%
  \raise.5ex\hbox{\the\scriptfont0 0}%
  \kern-.15em/\kern-.15em%
  \lower.25ex\hbox{\the\scriptfont0 00}}

\usepackage[colorlinks, linkcolor=black, citecolor=blue, urlcolor=blue, linktocpage, pdfencoding=unicode]{hyperref}

\begin{document}
\setcounter{page}{0}
\thispagestyle{empty}

\parskip 3pt

\font\mini=cmr10 at 2pt

\begin{titlepage}
	~\vspace{1cm}
	\begin{center}

		{\LARGE \bf $\lambda \phi^4$ Theory II:  The Broken Phase
			\\[8pt]Beyond NNNN(NNNN)LO}

		\vspace{0.6cm}

		{\large
			Marco~Serone$^{a,b}$, Gabriele~Spada$^{a,c}$, and Giovanni~Villadoro$^{b}$}
		\\
		\vspace{.6cm}
		{\normalsize { \sl $^{a}$
				SISSA International School for Advanced Studies and INFN Trieste, \\
				Via Bonomea 265, 34136, Trieste, Italy }}

		\vspace{.3cm}
		{\normalsize { \sl $^{b}$ Abdus Salam International Centre for Theoretical Physics, \\
				Strada Costiera 11, 34151, Trieste, Italy}}

		\vspace{.3cm}
		{\normalsize { \sl $^{c}$ Laboratoire Kastler Brossel, ENS - Universit\'e PSL, CNRS, Sorbonne Universit\'e,\\ Coll\`ege de France,  24 Rue Lhomond, 75005 Paris, France}}

	\end{center}
	\vspace{.8cm}
	\begin{abstract}

		We extend the study of the two-dimensional euclidean $\phi^4$ theory initiated in ref.~\cite{Serone:2018gjo} to the $\mathbb Z_2$ broken phase.
		In particular, we compute in perturbation theory up to N$^4$LO in the quartic coupling the vacuum energy, the vacuum expectation value of $\phi$ and the mass gap of the theory.
		We determine the large order behavior of the perturbative series by finding the leading order finite action complex instanton configuration in the $\mathbb Z_2$ broken phase.
		Using an appropriate conformal mapping, we then Borel resum the perturbative series.
		Interestingly enough, the truncated perturbative series for the vacuum energy and the vacuum expectation value of the field  is reliable up to the critical coupling
		where a second order phase transition occurs, and breaks down around the transition for the mass gap.
		We compute the vacuum energy using also an alternative perturbative series, dubbed exact perturbation theory, that allows us  to effectively reach
		N$^8$LO in the quartic coupling. In this way we can access the strong  coupling region of the $\mathbb Z_2$ broken phase and test Chang duality
		by comparing the vacuum energies computed in three different descriptions of the same physical system. This result can also be considered as a
		confirmation of the Borel summability of the theory.
		Our results are in very good agreement (and with comparable or better precision) with those obtained by Hamiltonian truncation methods.
		We also discuss some subtleties related to the physical interpretation of the mass gap and provide evidence that the kink mass can be obtained by analytic continuation from the unbroken to the broken phase.
	\end{abstract}

\end{titlepage}

\tableofcontents

\section{Introduction}

The $\phi^4$ theory in two dimensions is a particularly simple, yet non integrable, theory.
The UV divergencies are minimal and in the IR, for a critical value of the coupling,  it flows to the two-dimensional (2d) Ising model, that is an exactly solvable conformal field theory \cite{Onsager:1943jn,Belavin:1984vu}.
It also features a simple but non trivial duality symmetry, Chang duality \cite{Chang:1976ek}. For these reasons the $\phi^4$ theory is an ideal laboratory to
possibly test new methods, or improve on old ones, for studying quantum field theories at strong coupling, such as  lattice simulations
\cite{Milsted:2013rxa,Bosetti:2015lsa,Bronzin:2018tqz,Kadoh:2018tis}, hamiltonian truncations  \cite{Rychkov:2014eea,Rychkov:2015vap,Bajnok:2015bgw,Burkardt:2016ffk,Anand:2017yij,Elias-Miro:2017xxf,Elias-Miro:2017tup,Fitzpatrick:2018ttk,Chabysheva:2018wxr,Fitzpatrick:2018xlz} or resummation of the perturbative series \cite{Baker:1976ff,Baker:1977hp,LeGuillou:1979ixc,Orlov:2000wn,Pelissetto:2015yha,Serone:2018gjo}.
In the context of resummations, a connection
has been found recently between the Lefschetz thimble decomposition of path integrals and Borel summability of perturbative series \cite{Serone:2017nmd,Serone:2016qog},
that allowed us to show the Borel summability of a broad class of Euclidean 2d and 3d scalar field theories \cite{Serone:2018gjo}. These include the 2d  $\phi^4$ theory 
with both $m^2>0$, known already to be Borel resummable for parametrically small couplings \cite{Eckmann}, and $m^2<0$. 
\begin{figure}[t!]
	\centering
	\includegraphics[scale=.4]{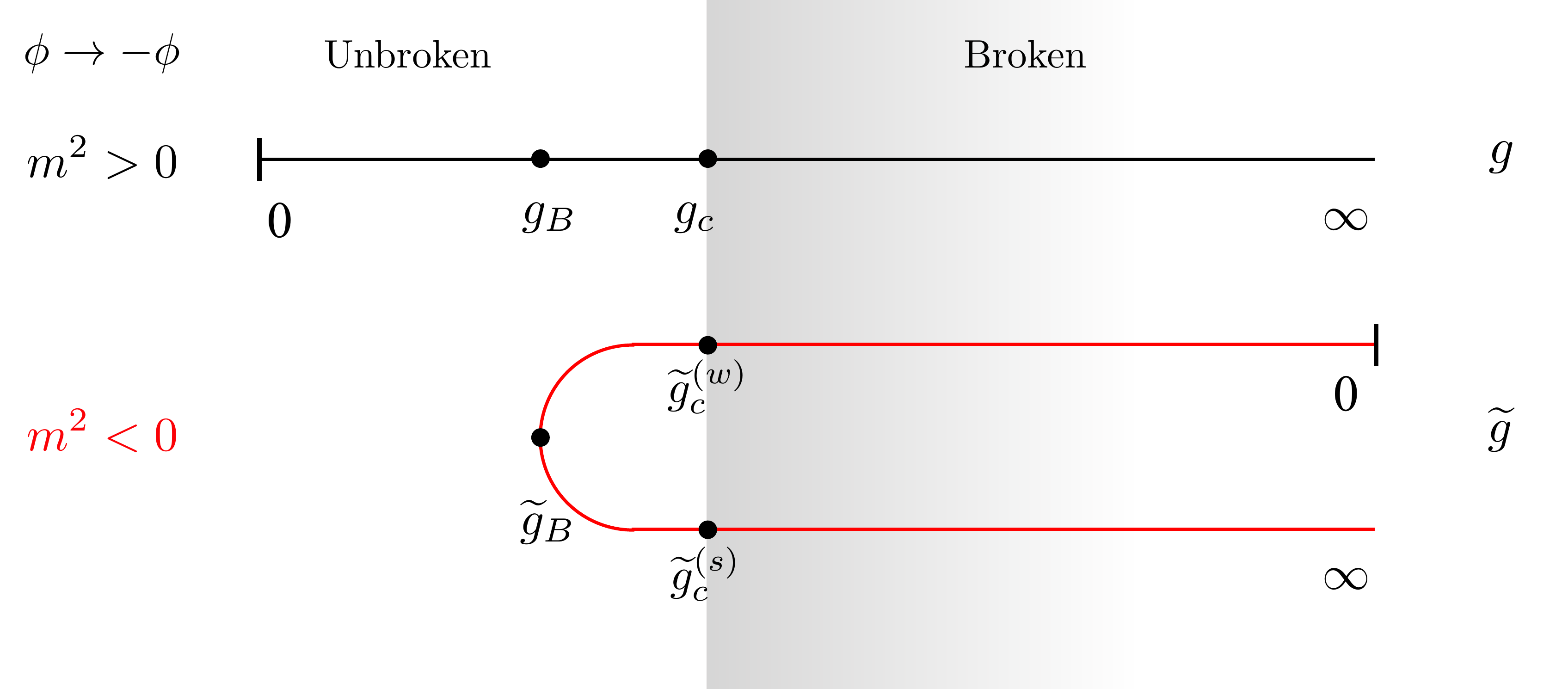}
	\caption{Phase structure of the $\phi^4$ theory according to Chang duality, as a function of the couplings $g$ and $\widetilde g$. Points in the same vertical line connecting the black and red lines correspond to
		different descriptions of the same theory.}
	\label{fig:ChangPic}
\end{figure}

The aim of this paper is to extend to the $\mathbb{Z}_2$ broken phase the study of the 2d Euclidean $\phi^4$ theory of ref.~\cite{Serone:2018gjo}, where the unbroken phase was analyzed.
In this way we will also provide numerical evidence of the Borel summability of the theory and the first applications of the methods introduced in refs.\cite{Serone:2017nmd,Serone:2016qog} in QFT.
The Euclidean Lagrangian density (modulo counterterms) reads
\begin{equation}
\Lagt = \frac 12 (\partial_\mu \phi)^2 - \frac 14 \widetilde m^2 \phi^2 + \lambda \phi^4 \,,
\label{Lagr}
\end{equation}
with $\widetilde m^2>0$. We denote the parameters in the broken phase with a tilde to distinguish them from the ones appearing in the Lagrangian density
\begin{equation}
\mathcal{L} = \frac 12 (\partial_\mu \phi)^2 + \frac 12 m^2 \phi^2 + \lambda \phi^4 \,, \quad g\equiv \frac{\lambda}{m^2}\,,
\end{equation}
used in ref.~\cite{Serone:2018gjo} to analyse the unbroken phase $(m^2>0)$.
The effective expansion parameter in the Lagrangian (\ref{Lagr}) is the dimensionless coupling
\be
\widetilde g \equiv \frac{\lambda}{\widetilde m^2}\,.
\ee
We start in section \ref{sec:ChangDual} by reviewing Chang duality and the phase structure of the theory, as expected by the duality.
This is summarized in fig.~\ref{fig:ChangPic}. We see that starting from a given value of the coupling, $g_B$ in the unbroken phase, the $\phi^4$ theory admits three equivalent descriptions:
one in terms of a theory with tree-level mass term $m^2>0$ and coupling $g = \lambda/m^2$ (black line) and two in terms of a theory with tree-level mass term $m^2 = - \widetilde m^2/2<0$ and coupling $\widetilde g$ (red lines).
We call weakly and strongly coupled branches the regions
$0\leq \widetilde g \leq \widetilde g_B$ and  $\widetilde g \geq \widetilde g_B$, respectively. The points $g_c$, $\widetilde g_c^{(w)}$ and $\widetilde g_c^{(s)}$  in fig.~\ref{fig:ChangPic} denote the critical couplings
in the three descriptions where the theory has a second order phase transition and flows to the 2d Ising model.
In the $\mathbb Z_2$ broken phase the only description that can be explored by resumming the perturbative series
and without encountering phase transitions is the part of the weakly coupled branch with  $0\leq \widetilde g \leq \widetilde g_c^{(w)}$. This is the region we will mostly focus on, although we will also explore other sectors of the phase diagram.
In ref.~\cite{Serone:2018gjo} we instead focused on the region $0\leq g \leq g_c$ in the black line of fig.~\ref{fig:ChangPic}.

In section \ref{sec:BorelGeom} we set the stage for the methods we will use in the paper to Borel resum the perturbative series.
In subsection \ref{subsec:NCM} we look for finite action solutions to the complexified Euclidean equations of motion (complex instanton configurations). As well-known \cite{Lipatov:1976ny},
the configurations with the smallest action determine the leading large-order behavior of the perturbative expansion of a given observable and the leading singularity
of its Borel transform function. This is useful information, that we exploit
to set up a suitable conformal mapping for the numerical Borel resummation of the perturbative series.
In subsection \ref{subsec:EPT} we consider a modified perturbative series, where the trilinear coupling defined below in eq.~(\ref{eq:g3def}) gets replaced by  $\lambda_3 \rightarrow  \sqrt{2} \lambda \mt^2/{\sqrt{\lambda_0}}$. It has been shown in refs.\cite{Serone:2017nmd,Serone:2016qog}, in the context of quantum mechanical models, that modified perturbative expansion
of this kind, where one expands in $\gt$ with $\gt_0\equiv \lambda_0/\widetilde m^2$ held fixed, and sets back $\gt_0=\gt$ {\it after} the resummation has been performed,
leads to a significant improvement of the Borel resummation of perturbative series at strong coupling. Expansions of this kind were dubbed in refs.\cite{Serone:2017nmd,Serone:2016qog} Exact Perturbation Theory (EPT).\footnote{In fact, tricks of this sort can do more than that, turning a non-Borel resummable ordinary expansion into a non-ordinary Borel-resummable one.}
EPT will allow us to compute the vacuum energy in the ${\mathbb Z}_2$ broken phase at strong coupling, namely in the region $\gt \geq \gt_0$ in fig.~\ref{fig:ChangPic}
(strongly coupled branch). Note that for $\gt\neq \gt_0$ the $\mathbb Z_2$ symmetry is explicitly broken, so EPT is equivalent to adding an explicit symmetry breaking term which is eventually set to zero.

In section~\ref{sec:PertCoeff} we compute the perturbative series expansion up to order $\gt^4$ of the 0-point, 1-point and 2-point Schwinger functions.
This is the maximal order that can be reached by computing Feynman diagrams with up to eight interaction vertices (the maximum we could reach), because of the presence
of the trilinear interaction proportional to
\be
\lambda_3 =\sqrt{2\lambda}\mt\,.
\label{eq:g3def}
\ee
We use a renormalization scheme which is equivalent to normal ordering, but perform our computations in an intermediate auxiliary scheme which allows us to efficiently treat the corrections to the 1-point tadpole
that are generated order by order in perturbation theory.  The main results of this section are the perturbative expressions for the vacuum energy $\widetilde \Lambda$, the 1-point function
$\langle \phi \rangle$ and the physical mass $\widetilde M^2$ in eqs.(\ref{eq:Lambda_series}), (\ref{eq:Tadpole_series}) and \eqref{eq:Mass_series}, respectively. In addition to that, we provide in eq.~\eqref{eq:Lambda_EPT} the perturbative expression of
$\widetilde \Lambda$ up to $\gt^8$ in EPT.

We report in section~\ref{sec:results} the final results of our investigation. We start by discussing the main differences in the numerical resummation procedures with respect to ref.~\cite{Serone:2018gjo}.
In particular, we point out that the shortness of the ordinary perturbative series, together with the slow convergence of the series after the conformal mapping,\footnote{Actually the series remains asymptotic after the conformal mapping because of other singularities in the Borel plane, but their effect is expected to be small in the region of interest.}
leads to results which are much less accurate than those found by Borel resumming in the unbroken case.
\begin{table}[t!]
	\centering
	{
		\renewcommand{\arraystretch}{0.8}
		\begin{tabular}{ccc}
			\toprule
			$\widetilde g_c^{(w)}$ & $g_c^{{\rm Chang}}$ & $g_c$       \\
			\midrule
			$0.29(2)$              & $2.64(11)$          & $2.807(34)$ \\
			\bottomrule
		\end{tabular}
	}
	\caption{Values of the critical coupling in the $\phi^4$ theory. The value $\widetilde g_c^{(w)}$ is the one found in this paper from eq.~(\ref{eq:vevIntro}),
	$g_c^{{\rm Chang}}$ is $\widetilde g_c^{(w)}$ expressed in terms of the unbroken variables using Chang relation (\ref{Chang2}) and $g_c$ is the value found in the unbroken phase in ref.\cite{Serone:2018gjo}.}
	\label{FinalRes}
\end{table}
On the other hand, we find that in the entire regime $0\leq \gt \leq \gt_c^{(w)}$, the ${\cal O}(\gt^4)$ perturbative series
is reliable for $\widetilde \Lambda$ and $\langle \phi \rangle$ while the series for $\widetilde M^2$ breaks down slightly before  $\gt_c^{(w)}$.
For all the three observables we find that the central values of the Borel resummed results are in very good agreement with the perturbative result, see figs.~\ref{fig:LambdaPT}, \ref{fig:tad} and \ref{fig:mass}.
We determine the critical coupling $\widetilde g_c^{(w)}$ by demanding
\be
\langle \phi \rangle|_{\widetilde g=\widetilde g_c^{(w)}} = 0\,.
\label{eq:vevIntro}
\ee
We report in table \ref{FinalRes} the value of $\widetilde g_c^{(w)}$. For convenience, we also report its value $g_c^{{\rm Chang}}$ in terms of the unbroken variables, using Chang duality,
and compare it with the value $g_c$ found in ref.\cite{Serone:2018gjo}. Chang duality predicts $g_c^{{\rm Chang}}=g_c$, compatible with our results.
As discussed in ref.~\cite{Serone:2018gjo}, the Borel resummed Schwinger functions are expected to coincide with the exact ones in a given phase of the theory.
At $\gt= \gt_c^{(w)}$  the two vacua of the broken phase collide in the unique $\mathbb Z_2$ invariant vacuum and
for $ \gt_c^{(w)}< \gt < \gt_c^{(s)}$ the $\mathbb Z_2$ symmetry is restored.
Consequently our perturbative results for $\langle \phi \rangle$ and $\widetilde M^2$ for $\gt > \gt_c^{(w)}$ do not have an obvious direct physical interpretation.\footnote{In fact we do not understand how to physically interpret $\widetilde M^2$ for  $\gt > \gt_{k\bar k}$, where $  \gt_{k\bar k} < \gt_c^{(w)}$, see section \ref{sec:comparisons}.}  On the other hand, $\widetilde \Lambda$ should be continuous along the transition and its value past $\gt_c^{(w)}$ should still be identified with the vacuum energy in the symmetric phase and then after $\gt_c^{(s)}$  with the vacuum energy in the strongly coupled branch of the broken phase. This comment applies also for the vacuum energy $\Lambda$ for $g> g_c$ computed starting from the unbroken phase with $m^2>0$.
We then compute $\widetilde \Lambda$ using the ${\cal O}(\gt^8)$ series associated to EPT, which allows us to explore the strongly coupled branch of the theory with a better accuracy than the
ordinary ${\cal O}(\gt^4)$ series. We determine $\widetilde \Lambda$ in the broken phase for a certain range of $\gt$ in all three descriptions, see fig.~\ref{fig:LambdaChang}.
We consider the agreement of the results as a convincing numerical check of Chang duality,\footnote{Chang duality has been numerically tested in ref.~\cite{Rychkov:2015vap} by comparing the vacuum energy in the unbroken and the weakly coupled branch of the broken phase, but no analysis of the strongly coupled branch was performed.} of the Borel summability of the theory and of the usefulness of EPT in QFT.

The vacuum energy and the mass gap of the 2d $\phi^4$ theory in the broken phase have also been computed using Hamiltonian truncation methods in refs.\cite{Rychkov:2015vap,Bajnok:2015bgw}.
We compare our findings with those of the above works in section \ref{sec:comparisons}, finding very good agreement. In particular, we confirm that both the results for $\widetilde \Lambda$ of ref.~\cite{Rychkov:2015vap} and the ones for $\widetilde M^2$ in ref.\cite{Bajnok:2015bgw} are within the regime of validity of perturbation theory!
The $\phi^4$ theory considered in refs.\cite{Rychkov:2015vap,Bajnok:2015bgw} is not however in a genuinely broken phase. This makes a comparison with refs.~\cite{Rychkov:2015vap,Bajnok:2015bgw} non-trivial in a range of the coupling where the elementary particle is supposed to decay in pair of kink and anti-kink, since single  kink states decouple in a theory where the $\mathbb Z_2$ symmetry is spontaneously broken.
In subsection \ref{subsec:masskink} we provide numerical evidence that the kink sector of the theory can be accessed starting from the unbroken phase considered in ref.~\cite{Serone:2018gjo}.
More precisely, we show that the value of the mass gap $|M|$  computed in the unbroken phase  for $g\geq g_c$ is in agreement with the mass of the kink state computed in ref.~\cite{Bajnok:2015bgw},
see fig.~\ref{fig:mkink}.

We conclude in section~\ref{sec:conclu}.
In appendix \ref{sec:appendix} the coefficients for the series expansion of the $0$, $1$ and $2$-point function are reported.
The coefficients have been determined for independent cubic and quartic couplings $\lambda_3$ and $\lambda$ and could be used also for theories with explicit breaking of the $\mathbb Z_2$ symmetry
$\phi\rightarrow - \phi$.
We would finally like to emphasize that while several works along more than forty years have analyzed the 2d $\phi^4$ theory in the unbroken case  by means of various resummation procedures \cite{Baker:1976ff,Baker:1977hp,LeGuillou:1979ixc,Orlov:2000wn,Pelissetto:2015yha,LeGuillou:1985pg,Kompaniets:2017yct,Serone:2018gjo},
as far as we know this is the first paper addressing the $\mathbb{Z}_2$ broken phase.

\section{Chang Duality}
\label{sec:ChangDual}

In the $\phi^4$ theory with $m^2>0$, aside from the normalization of the free theory path integral, there are only two superficially divergent one-particle irreducible (1PI) diagrams:
the two loop number eight graph occurring in the $0$-point function and the one-loop tadpole of the $2$-point function. Correspondingly, the counterterms $\delta \Lambda$ and $\delta m^2$
contain up to ${\cal O}(\lambda)$ terms to all orders in perturbation theory, with no need of a wave function and coupling constant counterterms.
Such simple renormalization property are at the base of one of the simplest strong-weak dualities in QFT: Chang duality \cite{Chang:1976ek}.
The original derivation worked with normal ordering prescriptions (recently nicely reviewed in ref.~\cite{Rychkov:2015vap}), but the same analysis can be repeated in other regularizations.
We choose here dimensional regularization (DR). This formulation allows for a straightforward generalization of the duality in the 3d $\lambda \phi^4$ theory \cite{Magruder:1976px}, where normal ordering is no longer enough to cancel all divergencies. The theory in $d$ dimensions defined by the (Euclidean) Lagrangian
\begin{equation}
	{\cal L}= \frac 12 (\partial\phi)^2 + \frac 12 m^2 \phi^2 + \lambda \mu^\epsilon \phi^4 + \frac 12 \delta m^2 \phi^2 + \mu^{-\epsilon} \delta \Lambda\,,
	\label{eq:LagI}
\end{equation}
where $\epsilon = 2 - d$ and $\mu$ is the usual RG sliding scale, has a dual description in terms of the theory $\Lagt$ with negative squared-mass term
\begin{equation}
	\Lagt = \frac 12 (\partial\phi)^2 - \frac 14 \mt^2 \phi^2 + \lambda \mu^\epsilon \phi^4 + \frac 12 \delta \mt^2 \phi^2 + \mu^{-\epsilon}\delta \Lt + \Delta \Lt \,,
	\label{eq:LagII}
\end{equation}
which is derived below using dimensional regularization and a modified Minimal Subtraction (MS) renormalization scheme.
The counterterms for $\cal L$ are fixed by requiring that for $\mu = m$ the 2-point tadpole diagram at one loop is exactly canceled and that the vacuum energy vanishes up to $\mathcal{O}(\lambda^2)$.\footnote{The counterterm $\delta\Lambda$ at the scale $\mu=m$ exactly cancels the free theory contribution proportional to $\int d^dp \log (p^2+m^2)$ as well as the $\mathcal{O} (\lambda)$ contributions from the two-loop ``8"-shaped diagram and the one-loop correction proportional to $\delta m^2$.} We get
\begin{gather}
	\delta m^2
	= -\frac{3\lambda}{\pi} \Big(\frac{2}{\epsilon} -\bar \gamma \Big) \,,
	\nn
	\\
	\delta \Lambda
	= \frac{3 \lambda }{4\pi ^2 \epsilon ^2}
	-\frac{1}{4 \pi \epsilon }\left( \frac{3 \lambda}{\pi} \bar \gamma + m^2(\mu)\right)
	-\frac{m^2}{8 \pi }
	+\frac{m^2 \bar \gamma}{8 \pi }
	+\frac{3 \lambda  {\bar \gamma}^2}{16 \pi ^2} \,,
	\label{eq:deltamLambda}
\end{gather}
where we defined $\bar \gamma \equiv \gamma_E - \log(4\pi)$, with $\gamma_E \approx 0.577\dots$ the Euler-Mascheroni constant. Note that in the above expression the $1/\epsilon$ term of $\delta \Lambda$ depends on the squared mass defined at the scale $\mu$ at which we are evaluating the theory (whose explicit expression is obtained below). This is needed for the counterterm to cancel the divergent part at any scale.
The finite contribution to $\delta \Lambda$ depends instead on the arbitrary renormalization point chosen, i.e.~$m^2$, and hence the squared mass terms that appear in the $\mathcal{O}(\epsilon^0)$ term are defined as $m^2(m)\equiv m^2$.
To all orders in perturbation theory the $\beta$ functions for the mass, the coupling and the vacuum energy read
\begin{equation}
	\beta_{m^2} = -\frac{6}{\pi}\lambda  \,,
	\quad \quad
	\beta_\lambda =0\,,
	\quad \quad
	\beta_\Lambda =
	-\frac{1}{4 \pi }\left( \frac{3 \lambda}{\pi} \bar \gamma + m^2(\mu)\right)
	\,,
\end{equation}
and hence
\begin{gather}
	m^2(\mu) = m^2 + \frac{3 \lambda}{\pi} \log \frac{m^2}{\mu^2}\,,
	\nn
	\\
	\Lambda(\mu) = \Lambda(m)
	+\frac{3 \lambda}{16\pi^2}  \log^2 \frac{m^2}{\mu^2}
	+\left(\frac{3 \lambda}{\pi } \bar \gamma + m^2 \right) \frac{1}{8\pi} \log \frac{m^2}{\mu^2}
	\label{eq:running}
	\,.
\end{gather}
Renormalizing with a normal ordering mass $\mu$ is equivalent to use $m^2(\mu)$ in the Lagrangian (\ref{eq:LagI}).
The counterterms for $\Lagt$ are chosen in the same modified MS renormalization scheme as before, i.e.~by requiring that for $\mu = \mt$ the 2-point tadpole diagram at one loop and the divergent 0-point terms are exactly canceled.  We obtain\footnote{The counterterm $\delta \Lt$ has a different form w.r.t.~$\delta \Lambda$ because it gets shifted by the quantity $\mt^2 \delta \mt^2/(16 \lambda)$ when the Lagrangian \eqref{eq:LagII} is expanded around the classical minimum.}
\begin{gather}
	\delta \mt^2
	= -\frac{3\lambda}{\pi} \Big(\frac{2}{\epsilon} -\bar \gamma \Big) \,,
	\nn
	\\
	\delta \Lt
	=
	\frac{3 \lambda }{4 \pi ^2 \epsilon ^2}
	-\frac{1}{4 \pi \epsilon } \left(\frac{3\lambda}{\pi} \bar \gamma   -\frac{1}{2}  \mt^2(\mu)\right)
	-\frac{\mt^2}{8 \pi }
	-\frac{\mt^2 \bar \gamma}{16 \pi }
	+\frac{3 \lambda  {\bar \gamma}^2}{16 \pi ^2} \,.
	\label{eq:counterterms}
\end{gather}
Using eq.~\eqref{eq:running} and setting $\mu=\mt$, we find that the theory \eqref{eq:LagI} is equivalent to \eqref{eq:LagII}
provided the following identities hold
\begin{gather}
	m^2 +\frac{3\lambda }{\pi}  \log \frac{m^2}{\mt^2} = -\frac 12 \mt^2
	\,, \nn \\
	\Delta \Lt =
	\frac{\mt^2-m^2}{8 \pi} + \frac{m^2}{8 \pi} \log \frac{m^2}{\mt^2}
	+\frac{3 \lambda}{16 \pi^2} \log^2\frac{m^2}{\mt^2}
	\,.
	\label{eq:Chang}
\end{gather}
We have then established a relation between two theories in different phases: the theory in the broken phase with squared mass term
$-\mt^2/4$ (giving rise to a state of mass squared $\mt^2$ when expanded around the tree-level VEV) in the modified MS scheme with $\mu=\mt$ is identical to the theory in the unbroken phase in the same
analogous modified MS scheme with $\mu=m$. This is Chang duality \cite{Chang:1976ek}.
The solutions to the Chang equation \eqref{eq:Chang} have been already reviewed in ref.~\cite{Rychkov:2015vap}, but for completeness we will again briefly summarize them here.
Defining the dimensionless coupling constants
\be
\gt = \frac{\lambda}{\mt^2} \,, \quad \quad  g = \frac{\lambda}{m^2} \,,
\ee
we can rewrite eq.\eqref{eq:Chang} as
\be
f(g) = \widetilde f(\gt)\,,
\label{Chang2}
\ee
where
\be
f(g) = \log g -\frac{\pi}{3g} \,, \quad  \widetilde f(\gt) =  \log \gt +\frac{\pi}{6\gt}\,.
\ee
At fixed $g$, we look for solutions in $\gt$ of eq.~(\ref{Chang2}). Since $ \widetilde f(\gt)>0$ $\forall \gt$, no solutions can evidently exist for sufficiently small $g$, where
$f(g)<0$. The minimum of $\widetilde f(\gt)$ occurs at $\gt_B = \pi/6$ and a solution exists for
\be
g \geq g_B = \frac{\pi}{3W(2/e)} \approx 2.26\,,
\ee
where $W$ is the Lambert function (also known as product logarithm or omega function). For $g \geq g_B$ there are two solution branches $\gt_{w,s}(g)$.
We label the two branches as strong ($s$) and weak ($w$) branches according to their behavior:
\be
\widetilde  g_{w}(g)\approx \frac{\pi}{6 \log  g} \,, \quad
\widetilde  g_{s}(g)\approx g\,, \quad  g\rightarrow \infty \,.
\ee
The existence of the weak branch allows to prove the existence of a phase transition in the $\phi^4$ theory in the classically unbroken phase with $m^2>0$ at sufficiently strong coupling
$g$. Indeed, for parametrically small $g$ the theory is well described by its classical potential and is in the unbroken phase, while at parametrically
large couplings the duality implies it is in the broken phase, since it can be described by a weakly coupled theory with $m^2=-\mt^2/2<0$. By continuity there should exist a critical coupling $g_c$
where the phase transition occurs. The value of the critical coupling in the normal ordering scheme has been computed by different methods \cite{Serone:2018gjo,Milsted:2013rxa,Bosetti:2015lsa,Bronzin:2018tqz,Kadoh:2018tis,Bajnok:2015bgw,Elias-Miro:2017xxf,Elias-Miro:2017tup,Pelissetto:2015yha} to be $g_c\approx 2.76$.
Correspondingly, we can predict the value of the two critical couplings:
\be
\gt_c^{(w)}(g_c)\approx 0.27\,, \quad \quad \gt_c^{(s)}(g_c)\approx 1.24\,.
\label{eq:gcfromChang}
\ee
The phase structure of the theory, as predicted from Chang duality, is then the following.
Starting from a perturbative description in the unbroken phase, $m^2>0$, the theory develops a (second-order) phase transition at $g=g_c$
and above $g_c$ remains in the $\mathbb Z_2$ broken phase.
Starting from a perturbative description in the broken phase, $m^2 = -\mt^2/2<0$, we first encounter a phase transition at $\gt=\gt_c^{(w)}$,
the $\mathbb Z_2$ symmetry is restored for $\gt_c^{(w)}<\gt<\gt_c^{(s)}$, and at $\gt=\gt_c^{(s)}$ we have another phase transition.
For $\gt > \gt_c^{(s)}$ the theory remains in the broken phase. Chang duality predicts that the three regimes $g_c \leq g \leq \infty$, $0\leq \gt \leq \gt_c^{(w)}$,
and $ \gt_c^{(s)} \leq \gt\leq \infty$ are different descriptions of the same physical theory in the broken phase. Similarly, the three regimes $g_B \leq g \leq g_c$, $g_c^{(w)} \leq \gt \leq \gt_B$, and $ \gt_B \leq \gt\leq  \gt_c^{(s)}$ are different descriptions of the same physical theory in the unbroken phase. In particular the three critical points represent the very same transition in different descriptions. The region $0\leq g \leq g_B$ admits a single description in terms of the unbroken theory with $m^2>0$. See fig.~\ref{fig:ChangPic} for a summary.
In this paper, we will mostly focus on the weak branch regime $0\leq \gt \leq \gt_c^{(w)}$. However, by using different techniques, we will be able to compute
the vacuum energy in the broken phase in all the three descriptions.

\section{Borel Summability in the Broken Phase}
\label{sec:BorelGeom}

As discussed in ref.~\cite{Serone:2018gjo} the
$\phi^4$ theory is Borel resummable to the exact result also in the
broken phase, when the perturbative expansion is performed around either one of the two degenerate vacua at infinite volume. Indeed, thanks to Derrick's theorem~\cite{Derrick:1964ww}, the
non-trivial real saddles connecting
the two vacua have infinite action at infinite volume and do not obstruct the Borel transformation procedure. Besides, the vacuum selection, performed by introducing a fictitious $\mathbb  Z_2$ breaking parameter removed only after taking the infinite volume limit, decouples any contribution from the Borel resummable perturbative expansion around the `other' vacuum. The perturbative expansion around
one of the two $Z_2$-breaking vacua is therefore Borel resummable to the exact result of the symmetry
broken theory, similarly to the unbroken case (and with the same caveats associated to phase transitions and manipulations of the $N$-point functions).

Borel summability implies that a generic observable\footnote{By a generic observable we mean
	a generic Euclidean $N$-point function and, to some extent, simple quantities derived from $N$-point functions such as vacuum energy and mass gap. } $F(\gt)$ function of the coupling $\gt$ (with a divergent series expansion $\sum_{n=0}^\infty F_n \gt^n$) can be recovered by performing the integral
\begin{equation} \label{eq:borel}
	F(\gt)=\frac{1}{\gt}\int_0^\infty dt\, e^{-t/\gt}{\cal B}(t)\,,
\end{equation}
where the Borel function ${\cal B}(t)$ is the analytic continuation of the function defined by the
convergent series $\sum_n F_n t^n/n!$. While Borel summability corresponds to ${\cal B}(t)$ being regular over the positive real axis, singularities (in general an infinite number of them) are
present in the complex $t$ plane.  Their position corresponds to the value of the action on all possible (complex) solutions of the classical equations of motion. The saddles with the minimum absolute value for the action determine the radius of convergence of the perturbative series of ${\cal B}(t)$, thus the leading  growth of the  perturbative coefficients: if we call $t_1^{\pm}$  the position of such saddles (real action complex saddles come in pair, $t_i^-=t_i^+{}^*$)
the leading asymptotic growth
of the coefficients is of the form $F_n\propto a^n \Gamma(n+b +1)+ h.c.$ (with $a\equiv 1/t_1^+$ and $b$ a constant). Because of its finite radius of convergence, the Borel function can be well approximated by
a truncated perturbative expansion only in the interval $0<t<|t_1^\pm|$, this determines
an intrinsic error in the reconstruction of $F(\gt)$ from eq.~(\ref{eq:borel}),
which at small coupling is  ${\cal O}(e^{-|t_1^\pm|/\gt})$, reproducing the accuracy limit of the original divergent asymptotic series for $F(\gt)$. In order to really improve over the truncated series, ${\cal B}(t)$ must be analytically continued beyond its radius of convergence $|t_1^\pm|$. Two commonly used techniques to achieve this are the method of Pad\`e approximants
and the conformal mapping. To be effective, the first method in general requires the knowledge of a large number of coefficients, unfortunately this is not our case so the
improvement obtained with this method is limited. The second technique exploits the knowledge of the position
of the singularities in the complex $t$ plane to perform a clever change of variable $t(u)$ after
which the integration in eq.~(\ref{eq:borel}) is mapped over the interval $u \in [0,1)$ and all the singularities of ${\cal B}$ on the unit circle $|u|=1$ of the complex $u$-plane:
\begin{equation} \label{eq:borelmap}
	F(\gt)=\frac{1}{\gt} \int_0^1 du\, \left |t'(u)\right | e^{-t(u)/\gt}\, {\cal B}[t(u)]\,.
\end{equation}
The new series expansion of $\widetilde {\cal B}(u)\equiv {\cal B}[t(u)]$
in powers of $u$ is therefore convergent over the entire range of integration---the
original divergent truncated series has been transformed into a convergent one!
Of course the method requires the knowledge of all singularities of ${\cal B}$, i.e. all the
finite action complex saddle points of the classical action. Except in very special cases
such information is not available. However, even the mapping of the sole leading
singularity $t_1^\pm$ to the unit circle in the $u$-plane represents a big improvement:
it effectively enlarges the radius of convergence of the original Borel function to the next-to-leading
singularities $t_2^\pm$ improving the accuracy of the series to ${\cal O}(e^{-|t_2^\pm|/\gt})$.
We discuss this technique more in detail in the next subsection, while in section~\ref{subsec:EPT}
we will describe a different approach based on EPT introduced in refs.~\cite{Serone:2017nmd,Serone:2016qog}.

\subsection{Weak Coupling: Conformal Mapping of Complex Saddles}
\label{subsec:NCM}

In order to implement the conformal mapping method we must find the saddle points of the classical action (at least the leading ones) and the required mapping $t=t(u)$. Saddles with finite action configurations must have the field flowing to the minimum of the potential at infinity in all Euclidean directions. We concentrate on $SO$(2) symmetric configurations which are expected to be the dominant ones. In polar coordinates the problem reduces to a system of 2nd order non-linear differential equations for the real and the imaginary part of the field as a function of the radial coordinate $r$.
We look for solutions with Neumann boundary conditions at $r=0$ and the trivial vacuum $\phi=v$ at $r=\infty$.

For the unbroken phase the problem could be simplified by focusing on purely imaginary solutions, since in such case the system collapses to a single differential equation corresponding to a simple shooting problem (see ref.~\cite{Brezin:1992sq}). The value of the position of the singularity in this case is real and negative ($t_1^{\rm u.}=-1.4626...$).

In the broken phase, non-trivial saddle points necessarily require both the real and the imaginary parts of the field to vary. With some work the complex trajectories for the field and the corresponding values of the action can be obtained by numerical integration, following the complex solutions from the unbroken case for increasing values of the $\phi^3$ deformation that convert the unbroken potential to the double-well one.
\begin{figure}[t]
	\centering
	\includegraphics[height=8cm]{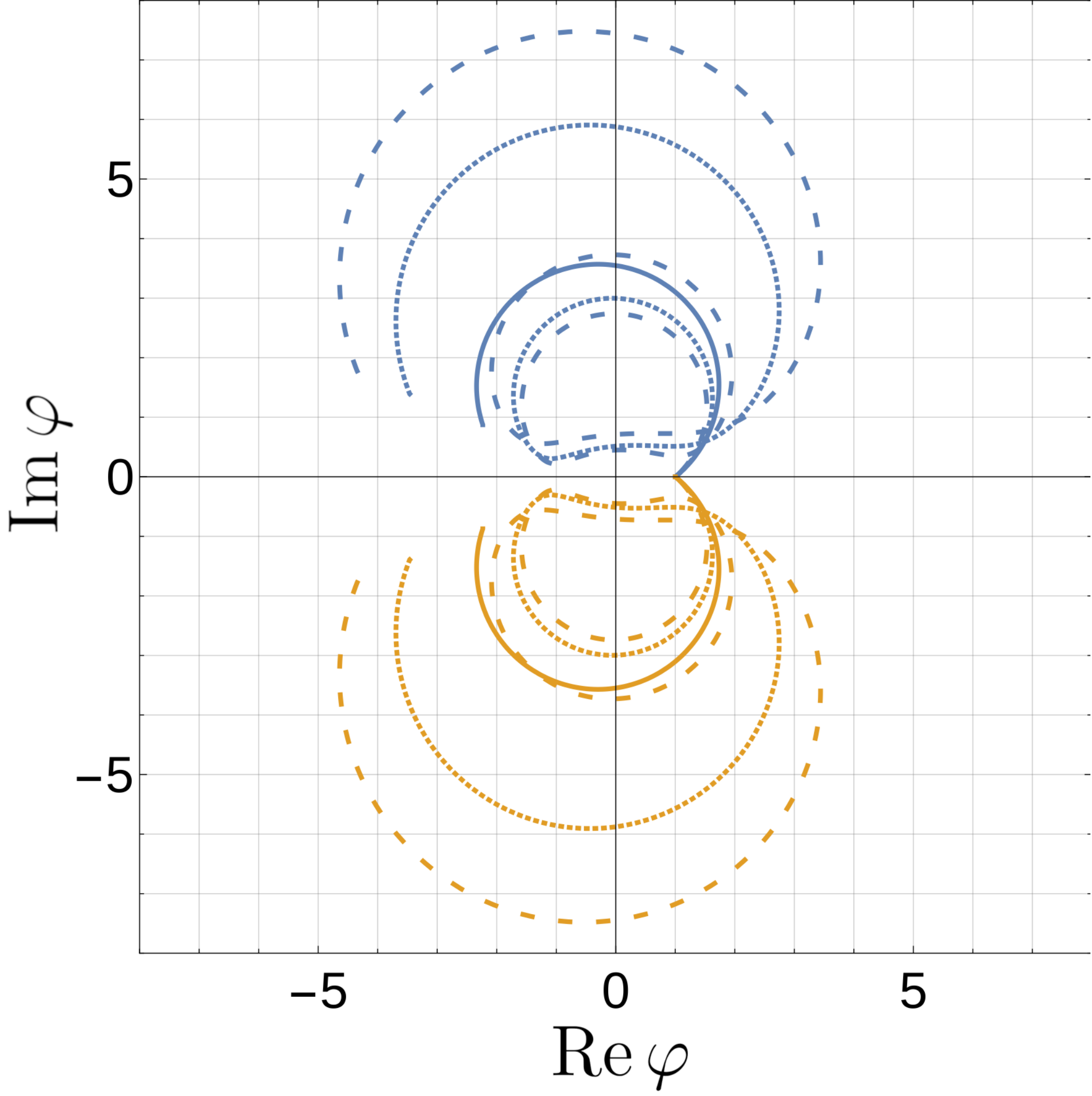}
	\caption{\label{fig:complexinst} Trajectories in the complex $\varphi$ plane of the solutions of eq.~\eqref{eq:radEOM}
		associated to the three leading saddles (blue) and their complex conjugates (orange). In increasing
		order of $|t_i|$ they are represented by continuous, dotted and dashed curves.
		All trajectories
		start for $\rho=0$ in three different points of the complex plane with zero velocities $\nabla \varphi(0)=0$
		and flow as $\rho \to \infty$ to the same value $\varphi(\infty)=1$. }
\end{figure}
After rescaling the field as $\phi = (8\gt)^{-1/2} \varphi$ and the coordinate as $r= \sqrt{2} \rho/\mt$ , the spherically symmetric solutions to the equation of motion satisfy the following differential equation
\begin{equation}
	\frac{\partial^2 \varphi}{\partial \rho^2}+\frac{1}{\rho} \frac{\partial \varphi}{\partial \rho} + \varphi - \varphi^3 =0 \,.
	\label{eq:radEOM}
\end{equation}
The trajectories for the first three leading
saddle points (among the $SO(2)$ invariant ones) are shown in fig.~\ref{fig:complexinst}.
They always come in complex conjugate pairs $\varphi_i^{\pm}$ and are quite non-trivial.
Interestingly enough, the complex trajectory of the leading saddle exactly matches
a circular arc in the complex $\varphi$-plane.
The corresponding values for the action
computed numerically lead to
the following values for the position of the leading singularities in the complex $t$ plane:
\begin{equation}
	\label{eq:tipm}
	t_1^\pm\simeq 1.10779544\pm 1.17944690\, i \,, \qquad t_2^\pm\simeq 8.64\pm 2.12\,i \,, \qquad t_3^\pm \simeq 22.8\pm 3.1\, i\,,
\end{equation}
where the numerical error should be smaller than the last digit reported, where
\be
t_i^\pm \equiv \frac{1}{\gt} \Big( S(\phi=\sqrt{8\gt} \, \varphi_i^\pm)-S(\phi=v)\Big) \,.
\ee
As mentioned before the value of the leading singularities determines the leading growth of the
perturbative coefficients. In particular $|t_1^\pm|=1.6181...$ is the radius of convergence of the Borel transform and roughly measures for what value of the coupling the theory turns completely non-perturbative. The fact that the second saddle is far at $|t_2^\pm|=8.89...$ means that,  for our purposes, it can be neglected in the conformal mapping. On the other hand $\pi/arg(t_1^+)\simeq3.8$ measures the half-period of oscillation of the sign of
the coefficients, in particular the vicinity of the saddle to the real axis decreases the oscillation of the  perturbative coefficients affecting its convergence.\footnote{This is also true
	for resummation methods like Pad\`e(-Borel) which rely on the oscillation of the coefficients to correctly reconstruct the position of the singularities.} We do not know the position (if present)
of $SO(2)$ non-invariant complex saddles, however they are expected to be subleading with respect to
the leading $SO(2)$ invariant one.

\begin{figure}[t!]
	\centering
	\includegraphics[width=.35\linewidth]{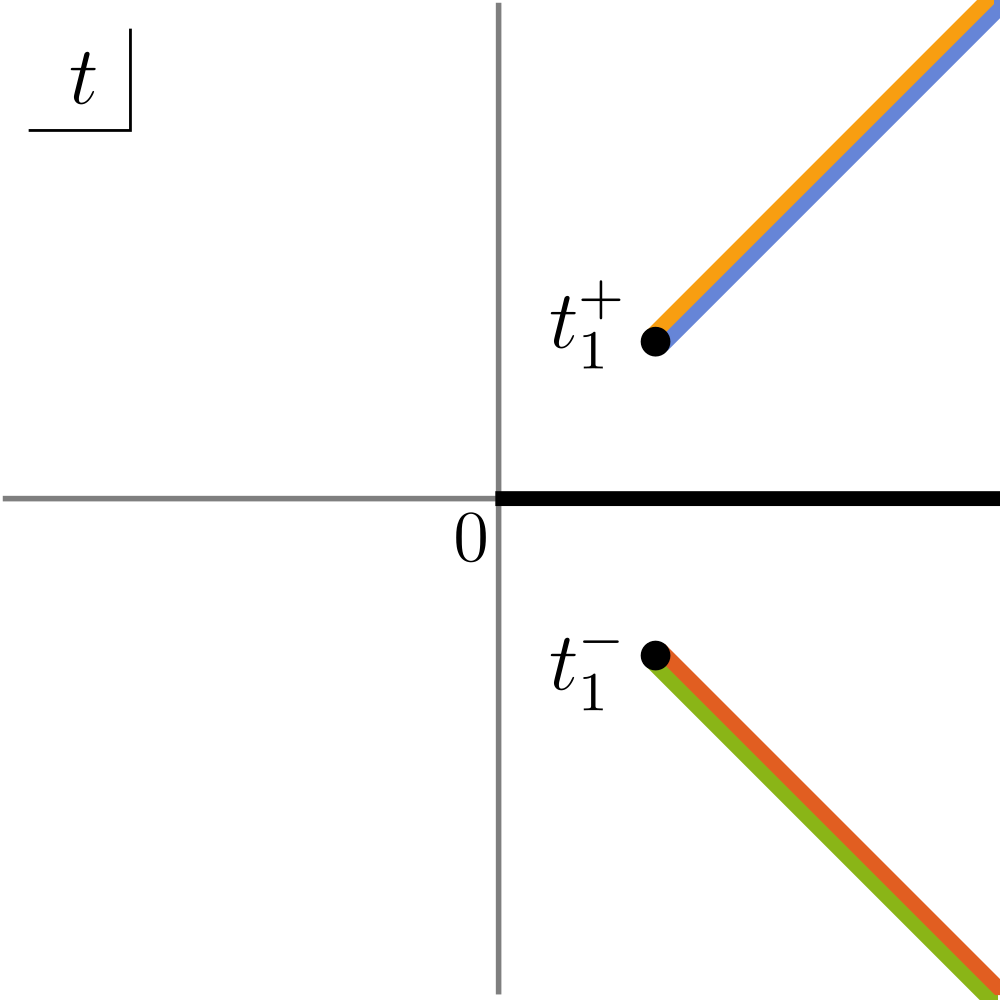}
	\hspace*{30pt}
	\hspace*{30pt}
	\includegraphics[width=.35\linewidth]{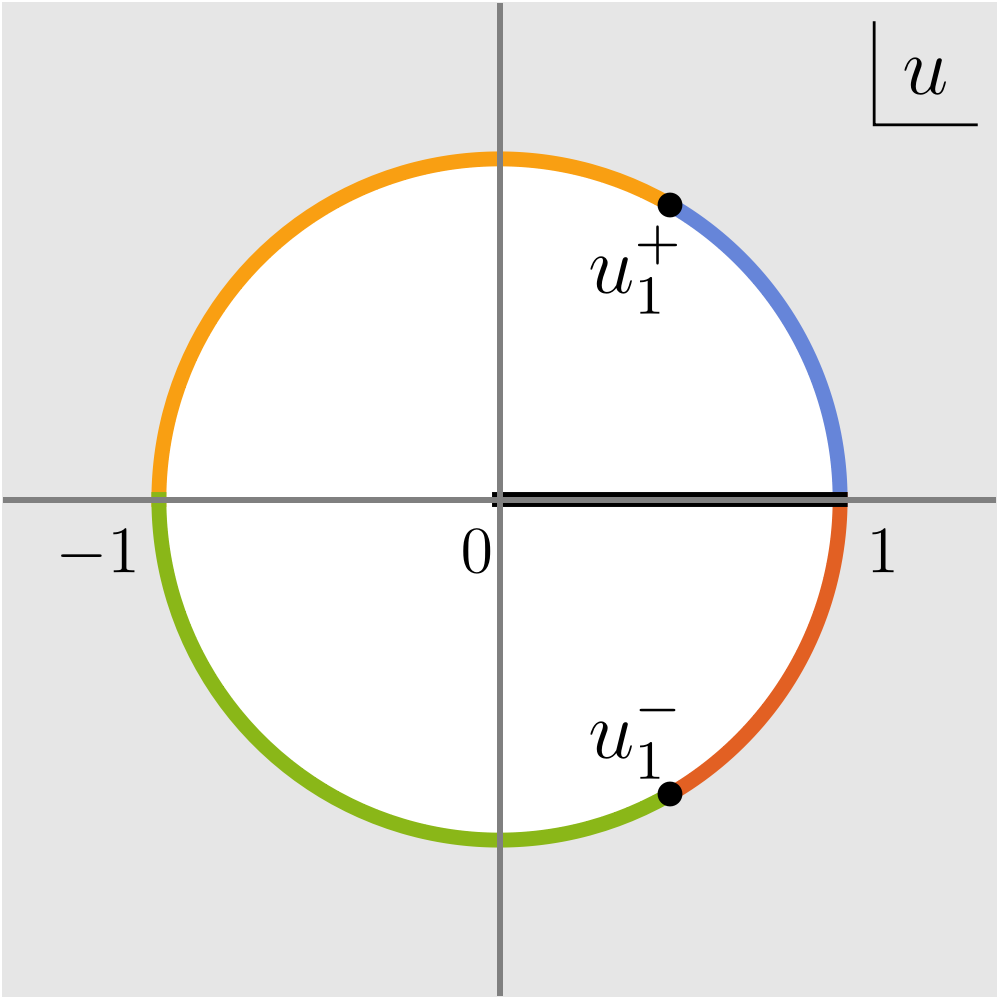}
	\caption{\label{fig:CM-mapgen} The change of variable~\eqref{eq:compmap} maps the cut $t$-plane (left panel) into the disk of unit radius $|u|=1$ (right panel). The branch points $t_1^\pm$ are mapped into the points $u_1^\pm$; the real positive axis $t \in [0,+\infty)$ is mapped in the segment $u \in [0,1)$; the rays connecting the branch points $t_1^\pm$ to the point at complex infinity are mapped to the arcs at the boundary of the disk as shown by the colors. In drawing this figure we picked $\alpha_1 = 1/4$.}
\end{figure}
Having identified the position of the leading singularities we now turn to the identification of
the suitable change of variable $t=t(u)$ for the resummation. This should correspond to the
(conformal) map which sends the positive semi axis $t\in [0,\infty)$ to the segment $u\in [0,1)$,
all singularities $t_i^\pm$ to the unit circle $|u|=1$ and be regular around the origin $t=u=0$. Schwarz-Christoffel transformations, which can map (degenerate) polygons to the unit disc,
exactly achieve this.  For the simple case of the mapping of a single couple of complex singularities located at the complex conjugate points $t=|t_1^\pm|e^{\pm i \pi \alpha_1}$ (with $\alpha_1 \in (0,1]$) the mapping takes the simple form\footnote{See ref.~\cite{Rossi:2018urw} for an application of the conformal mapping (\ref{eq:compmap}) in Borel resummation  methods.}
	\begin{equation} \label{eq:compmap}
		t=4 |t_1^\pm| \, u \left[\frac{\alpha_1}{(1-u)^2}\right]^{\alpha_1}
		\left[\frac{1-\alpha_1}{(1+u)^2}\right]^{1-\alpha_1}.
	\end{equation}
	Notice that for a real negative singularity ($\alpha_1=1$)
	the usual conformal mapping, used also in the unbroken
	phase (see e.g. ref.~\cite{Serone:2018gjo}), is recovered. For generic complex instantons
	eq.~(\ref{eq:compmap}) maps the singularities and the associated rays on the unit circle $|u|=1$ as in fig.~\ref{fig:CM-mapgen}.
	\begin{figure}[t]
		\centering
		\includegraphics[width=7cm]{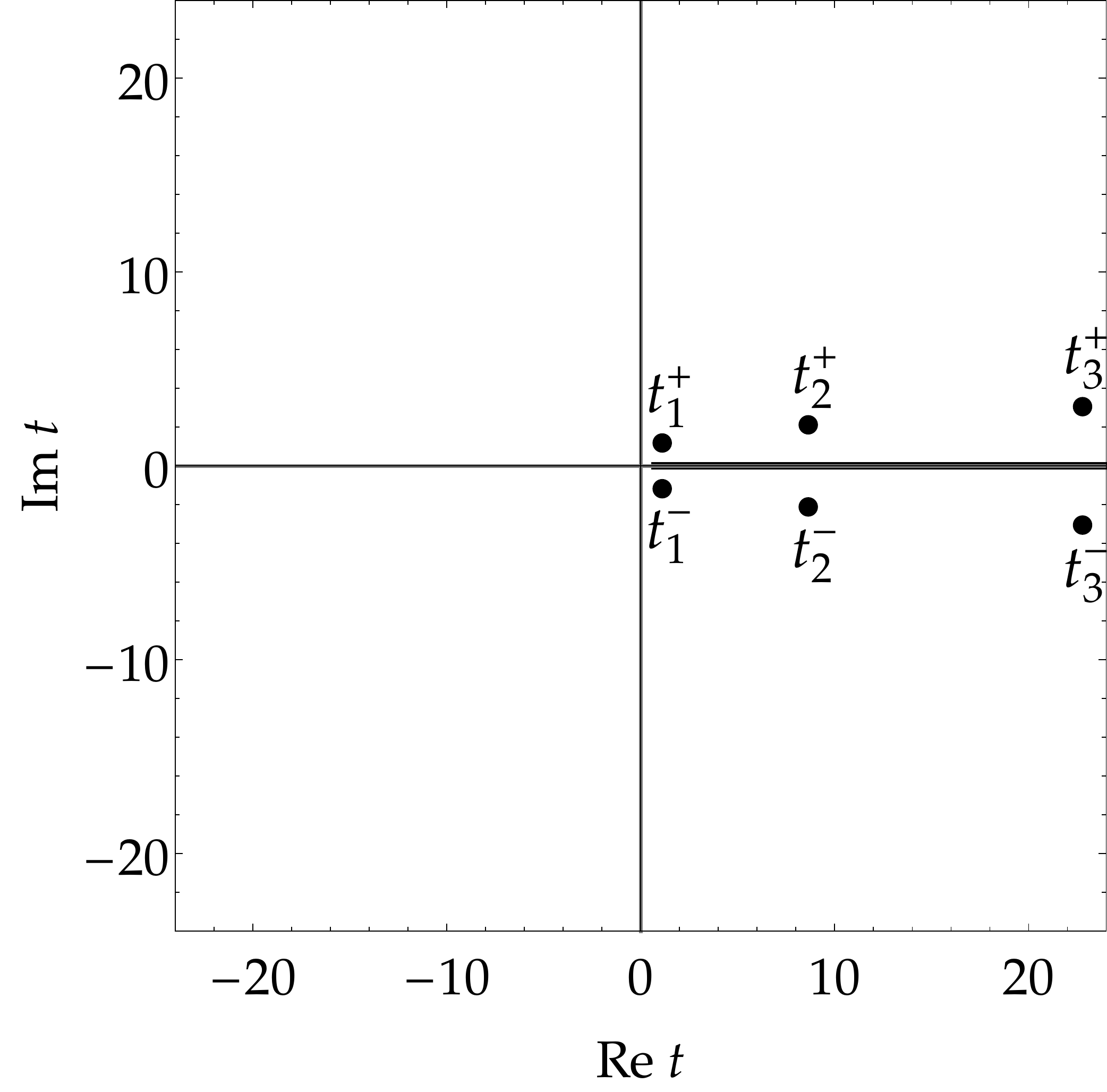} \qquad
		\includegraphics[width=7cm]{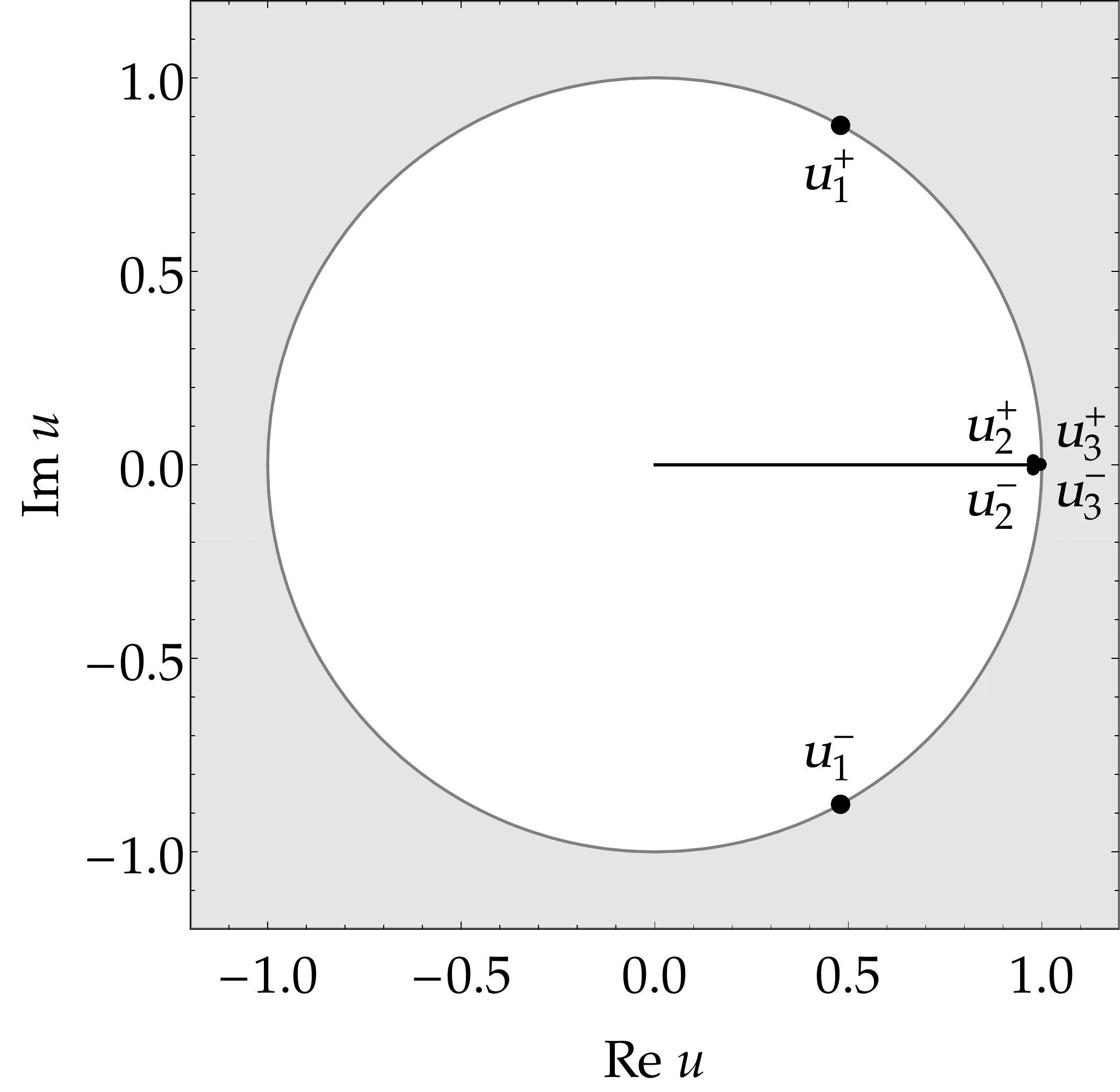}
		\caption{\label{fig:singtu} Position of the three leading singularities of ${\cal B}(t)$
			in the $t$ plane (left) and in the $u$-plane (right) after the conformal mapping of eq.~(\ref{eq:compmap}). The singularities $u_2^\pm$ and $u_3^\pm$ do not lie  at the unit circle $|u|=1$, though they are very close to it.}
	\end{figure}

	The factor
$e^{-t(u)/\gt}$ in eq.~(\ref{eq:borelmap})
	exponentially suppresses the integrand as $u\to1$. In the absence of
	any other singularity besides those mapped on the unit circle,
$\widetilde{\cal B}(u)$ has unit radius of
	convergence, and its truncated power series in $u$ can be used to approximate $\widetilde {\cal B}(u)$ with arbitrary precision on any point of the integration interval. Hence we can legitimately exchange the integration sign in eq.~(\ref{eq:borelmap}) with the summation sign of the power series defining $\widetilde {\cal B}(u)$. In this case the conformal mapping has turned the original asymptotic series in $\gt$ into a new convergent series for every finite value of $\gt$. This is not true in the presence of other singularities (as in our case) since the latter reduce the radius of convergence of
$\widetilde {\cal B}(u)$ to $|u_2^\pm|<1$
	(where $u_2^\pm$ is defined from $t(u_2^\pm)=t_2^\pm$) and the above exchange of integration and summation signs is no longer legitimate. The conformally mapped
	perturbative expansion can therefore approximate the integrand of eq.~(\ref{eq:borelmap}) only up to $|u_2^\pm|$, which at small coupling corresponds to an irreducible error of ${\cal O}(e^{-|t_2|/\gt})$. In this
	case the original series is mapped into another asymptotic one but with a smaller degree of divergence. Similarly to the hyperasymptotics techniques of refs.~\cite{HyperAs,HyperAs2} the knowledge
	of nearby complex instantons can be used to improve the degree of convergence of the original series.
	In our specific case the known subleading singularities $t_{i>1}^\pm$ are far enough (see fig.~\ref{fig:singtu}) and
	given the number of coefficients available and the range of couplings probed, the map (\ref{eq:compmap}) is not yet limited by the presence of these subleading instantons.

	While in principle any singularity can be moved away by an appropriate choice of $t(u)$,
	the efficiency of the mapping critically depends on the original position of the singularities,
	in particular on the
	angular distance from the positive real axis. For singularities on the negative real axis
	($\alpha=1$), the region of points at $t>|t_1^\pm|$ (which carry the non-trivial information
	required to improve beyond the original truncated expansion) is mapped into the region $u>3-2\sqrt2\simeq0.17$, well inside the radius of convergence of $\widetilde {\cal B}(u)$.
	As the singularities move closer to the positive real axis in the $t$-plane,
	such region is pushed more and more
	towards the unit circle in the $u$-plane,
	in particular for $\alpha_1\to 0$, the region with $t>|t_1^\pm|$ is
	squeezed into $u\gtrsim 1-\sqrt{\alpha_1}$. When the pair of complex singularities pinch the positive real axis the region $t>|t_1|$ becomes inaccessible and the series non-Borel summable.
	As a result the closer the singularities are to the real axis the less performant the conformal
	mapping is, and the larger is the number of coefficients requested to reach a certain accuracy.
	As we will discuss in section~\ref{sec:results} this is the main limiting factor of this method
	in our computation.

	Note that the conformal mapping (\ref{eq:compmap}) would have mapped the original asymptotic series into a convergent one if the
	other saddles were aligned among them (including the origin) along the cut $t$-plane depicted in fig.~\ref{fig:singtu}. This is the situation expected in the unbroken $Z_2$-symmetric phase, where the known singularities are aligned over the real negative $t$ axis and are mapped at the boundary of the unit $u$-disc.
	As we have seen, this is not the case in the spontaneously broken phase, where the saddles are not aligned.

	\subsection{Strong Coupling: Exact Perturbation Theory}
	\label{subsec:EPT}

	In the context of quantum mechanics it has recently been shown that it is possible to define modified perturbative expansions
	that are Borel resummable in theories where ordinary perturbation theory  is not \cite{Serone:2017nmd,Serone:2016qog}.
	For this reason such modified perturbative series were denoted Exact Perturbation Theory (EPT) in refs. \cite{Serone:2017nmd,Serone:2016qog}.
	EPT can also be useful when the ordinary expansion is Borel resummable to start with, by improving the Borel resummation of perturbative series. Namely, it can improve
	the accuracy of results obtained by the approximate knowledge of the Borel function with a finite number of perturbative terms, especially at strong coupling.
	We now show that EPT can similarly be applied in QFT. We will focus in what follows to the 2d $\phi^4$ theory, though most considerations apply also in more general settings.

	Consider a $n$-point Schwinger function
	\begin{equation}
		G^{(n)}(x_1,\ldots,x_n)={\cal N}  \int {\cal D}\phi\, \phi(x_1)\ldots \phi(x_{n})\, e^{-\int d^2 x \,\Lagt} \,,
		\label{eq:G0n}
	\end{equation}
	where ${\cal N}$ is an irrelevant constant factor, $\phi$ is the quantum fluctuation around the classical minimum $\phi_\mathrm{cl} = + \mt/\sqrt{8\lambda}$ and $\Lagt$ is the Lagrangian (\ref{eq:LagII}) expanded around $\phi_\mathrm{cl}$:
	\begin{equation}
		\Lagt= \frac 12 (\partial\phi)^2 + \frac 12 \mt^2 \phi^2 +\lambda_3  \phi^3 + \lambda \phi^4  \,,
		\label{sec:Lagt}
	\end{equation}
	where
	\be
	\lambda_3 =  \sqrt{2 \lambda} \mt\,.
	\ee
	For simplicity we have omitted to write in eq.~(\ref{sec:Lagt}) the counterterms $\delta \widetilde m^2$ and $\delta \widetilde \Lambda$ in eqs.(\ref{eq:counterterms}). The first is higher order in $\lambda$, while the last is field-independent,
	so they can be neglected when establishing the classical finite action field configurations. It is straightforward to see that the expansion in $\lambda$ is equivalent to a loopwise expansion in $\hbar$ and that all the terms
	appearing in the Lagrangian (\ref{sec:Lagt}) are of the same order in $\hbar$.
	Consider now the following $n$-point Schwinger function:
	\begin{equation}
		\hat G^{(n)}(x_1,\ldots,x_n,\lambda_0)=\hat {\cal N}  \int {\cal D}\phi\, \phi(x_1)\ldots \phi(x_{n})\, e^{-\int d^2 x \,\hat {\cal L}} \,,
	\end{equation}
	where
	\begin{equation}
		\hat {\cal L}= \frac 12 (\partial\phi)^2 + \frac 12 \mt^2 \phi^2 +  \hat \lambda_3\phi^3 + \lambda \phi^4  \,,
		\label{sec:Laghat}
	\end{equation}
	and
	\be
	\hat \lambda_3 = \mt  \lambda \sqrt{\frac{2}{\lambda_0}}\,.
	\ee
	For simplicity, as before we omit to write the counterterms necessary to make the theory UV finite.
	The Lagrangian (\ref{sec:Laghat}) is identical to that in eq.~(\ref{sec:Lagt}) except for the cubic coupling. At {\it fixed} $\lambda_0$,  we have effectively turned the classical cubic term into a quantum one.
	Hence the classical finite action configurations of  eq.~(\ref{sec:Laghat}) coincide with those
	of a theory in the $\mathbb Z_2$ unbroken phase with mass term $\widetilde m^2$ and quartic coupling $\lambda$, although of course the perturbative expansions in the two theories are different because of  the cubic term.
	At strong coupling the Borel resummability of the modified perturbative series of $\hat G^{(n)}$ is expected to be
	similar to the one of the unbroken  $\phi^4$ theory and hence better than that of the original expansion in $\lambda$ of $G^{(n)}$.
	We can then Borel resum the modified perturbative series of $\hat G^{(n)}$ and, after that, recover the original Schwinger function by setting $\lambda_0 = \lambda$:
	\be
	\hat G^{(n)}(x_1,\ldots,x_n,\lambda_0=\lambda)= G^{(n)}(x_1,\ldots,x_n)\,.
	\ee
	Note that whenever  $\lambda_0 \neq \lambda$ the Lagrangian $\hat {\cal L}$ breaks {\it explicitly} the  $\mathbb Z_2$ symmetry.
	We have then explicitly broken the symmetry,
	resummed, and switched off the breaking term after the resummation. This is precisely what we are supposed to do any time a vacuum should be non-perturbatively selected in presence of spontaneous symmetry breaking.
	While the Green functions $G^{(n)}$ should have singularities at the phase transition points,
	the Green functions $\hat G^{(n)}$ are smooth for $\lambda_0\neq \lambda$ and should become singular only in the limit $\lambda_0\rightarrow \lambda$.

	However, at weak coupling EPT is not as good as ordinary perturbation theory and requires more terms to correctly reproduce the weak coupling expansion of $G^{(n)}$.
	In particular,  we have verified that with the number of perturbative terms at our disposal, EPT can be reliably used only for the vacuum energy.

	\section{Perturbative Coefficients up to \texorpdfstring{$g_3^8$}{g\_3\^{8}}  Order}
	\label{sec:PertCoeff}

	In the 2d $\phi^4$ theory the vacuum energy and the mass are the only terms that require the introduction of counterterms
$\delta \Lambda$ and $\delta m^2$.
	In the broken phase, when we perform the vacuum selection by shifting the field $\phi \to \phi_\mathrm{cl}+ \phi$, a divergent one-loop 1-point tadpole is also generated (but the corresponding counterterm is fixed in terms of $\delta m^2$) as well as a cubic interaction term.
	Because of the cubic term in the Lagrangian, at a given order in the perturbative expansion the number of topologically distinct diagrams for the broken symmetry phase is much higher than for the symmetric phase. Moreover, diagrams with different number of cubic and quartic vertices contribute at each order in the effective coupling $\gt = \lambda/\mt^2$, making cancellations possible and lowering the numerical accuracy. The task of computing the perturbation series in this theory is then much more challenging.
	It should also be noted that, in contrast to the unbroken phase, this renormalization scheme is not optimal for computations because of the presence of a radiatively generated 1-point tadpole that should be taken into account order by order in perturbation theory.
	The computation of the $n$-point functions will in fact involve diagrams decorated with 1-point tadpole terms---i.e.~sub-diagrams with zero net momentum flow.
	A better renormalization scheme can be found by considering the auxiliary Lagrangian $\overline{\cal L}$ obtained by expanding eq.~\eqref{eq:LagII} around an arbitrary constant configuration $\phi \to \left(\phi_\mathrm{cl}+\phi_0\right) + \varphi$ and choosing the scale $\mu$ such that the divergent 2-pt tadpole is exactly canceled. Choosing $\phi_\mathrm{cl} = + \mt/\sqrt{8\lambda}$ and normalizing the vacuum energy so that $\Delta \Lt=0$ in eq.~\eqref{eq:LagII},  we get
	\begin{equation}
		\overline{\cal L}= \frac 12 (\partial\varphi)^2 +  \overline\lambda_1 \varphi + \frac 12 \overline m^2 \varphi^2 +  \overline\lambda_3 \varphi^3 + \lambda \varphi^4
		+ \Delta  \overline\Lambda
		+ c.t.
		\,,
		\label{eq:Lagh_aux}
	\end{equation}
	where we defined
	\begin{equation}
		\begin{aligned}
			\overline\lambda_1       & = 3 \lambda_3 \,Z +(\mt^2+ 12 \lambda Z )\phi_0 + 3 \lambda_3 \, \phi_0^2+ 4\lambda \phi_0^3
			\,,                                                                                                                                                                                                            \\
			\overline m^2            & = \mt^2 +12 \lambda Z +6\lambda_3 \, \phi_0 +12 \lambda \phi_0^2
			\,,                                                                                                                                                                                                            \\
			\overline\lambda_3       & = \lambda_3 + 4 \lambda \phi_0
			\,,                                                                                                                                                                                                            \\
			\Delta  \overline\Lambda & = \frac{\overline m^2-\mt^2}{8\pi} + \frac{\mt^2}{2}Z + 3 \lambda Z^2 +3 \lambda_3 \, Z \phi_0 + \frac 12 (\mt^2 +12 \lambda Z) \phi_0^2 + \lambda_3 \, \phi_0^3 + \lambda \phi_0^4
			\,,
		\end{aligned}
		\label{eq:param_aux}
	\end{equation}
	with $Z = (4\pi)^{-1}\log \left(\mt^2/ \overline m^2\right)$ and $\lambda_3 = \sqrt{2 \lambda} \mt$.
	The term $3 \lambda_3 Z$ for $ \overline\lambda_1$ and the term $3\lambda_3 Z \phi_0$ for $\Delta  \overline\Lambda$ come from the running of the linear term due to the one-loop tadpole divergence.
	The counterterms for the mass and the vacuum energy have the same form as in eq.~\eqref{eq:deltamLambda} with $m$ replaced by $ \overline m$ and the counterterm for the linear term is given by $\delta m^2  \overline\lambda_3/(4\lambda)$.
	In this theory all the divergent diagrams are exactly canceled and leave no finite part. If we additionally choose $\overline\lambda_1$ such that the 1PI one-point function $\widetilde \Gamma_1$ vanishes,\footnote{Following the notation of ref.\cite{Serone:2018gjo}, the tilde in $\Gamma_1$, and in $\Gamma_2$ in the following, refers to the Fourier transform of the 1PI correlation function.} the Lagrangian in eq.~\eqref{eq:Lagh_aux} is the most convenient for computations since we can now forget about both 1-point and 2-point tadpole terms.

	Practically the auxiliary theory $\overline {\cal L}$ can be used
	to determine the perturbative series in the 2d $\phi^4$ theory $\Lagt$ (in the renormalization scheme defined after eq.~\eqref{eq:running}) for arbitrary cubic coupling $\lambda_3$.
	We first fix the perturbative series for $\overline\lambda_1$ in terms of $\overline\lambda_3$ and $\lambda$ by requiring $\widetilde \Gamma_1=0$. Up to three loops the non-vanishing diagrams are
	\begin{equation}
		\begin{split}
			0 = \widetilde \Gamma_1 =& \,
			\overline\lambda_1
			+ \raisebox{-.5\height}{\includegraphics[width=.7cm]{figs/1pt-30}} \overline\lambda_3^3
			- \raisebox{-.5\height}{\includegraphics[width=.7cm]{figs/1pt-11}} \overline\lambda_3 \lambda
			+ \left[
				\raisebox{-.5\height}{\includegraphics[width=.7cm]{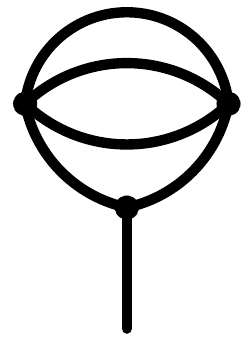}}
				+\raisebox{-.5\height}{\includegraphics[width=.7cm]{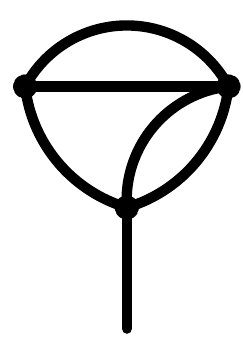}}
				\right]  \overline\lambda_3 \lambda^2
			+\left[
				\raisebox{-.5\height}{\includegraphics[width=.7cm]{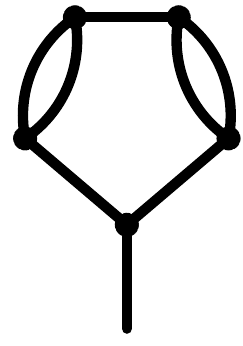}}
				+\raisebox{-.5\height}{\includegraphics[width=.7cm]{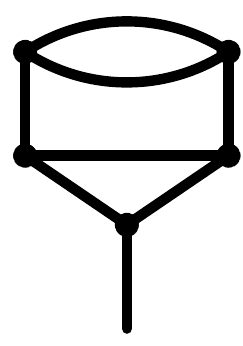}}
				+\raisebox{-.5\height}{\includegraphics[width=.7cm]{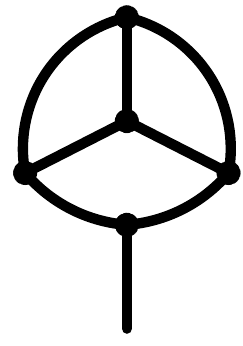}}
				\right]  \overline\lambda_3^5
			\\
			&
			-\left[\raisebox{-.5\height}{\includegraphics[width=.7cm]{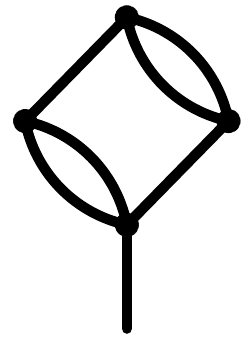}}
				+\raisebox{-.5\height}{\includegraphics[width=.7cm]{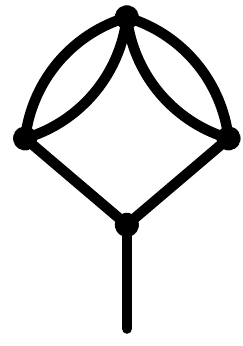}}
				+\raisebox{-.5\height}{\includegraphics[width=.7cm]{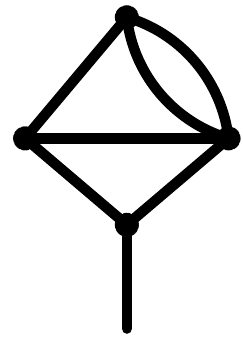}}
				+\raisebox{-.5\height}{\includegraphics[width=.7cm]{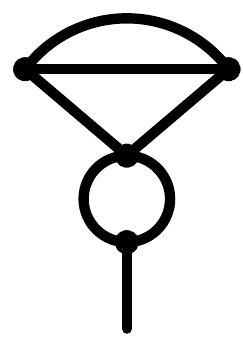}}
				+\raisebox{-.5\height}{\includegraphics[width=.7cm]{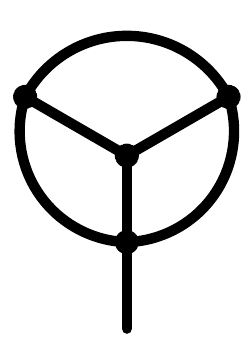}}
				\right]  \overline\lambda_3^3 \lambda + \dots
			\,,
		\end{split}
		\label{eq:G1_topo}
	\end{equation}
	where we have omitted the multiplicities. Therefore we get
	\begin{equation}
		\overline\lambda_1 =  c_{30} \frac{ \overline\lambda_3^3}{\overline m^4} + c_{11} \frac{\overline\lambda_3 \lambda}{{\overline m^2}} + c_{12} \frac{\overline\lambda_3 \lambda^2}{{\overline m^4}} + c_{50} \frac{\overline\lambda_3^5}{\overline m^8} + c_{31}\frac{\overline\lambda_3^3 \lambda}{\overline m^6}  + \dots
		\,,
	\end{equation}
	where $c_{ij}$ are the dimensionless coefficients obtained by computing all the Feynman diagrams with $i$ cubic and $j$ quartic vertices.
	We then use the set of eqs.~\eqref{eq:param_aux} to determine  $\phi_0$ and $\overline m^2$ as functions of $\mt$, $\lambda$ and $\lambda_3$. The first few orders are given by
	\begin{equation}
		\begin{aligned}
			\phi_0(\mt,\lambda,\lambda_3)
			 & = c_{11}\frac{\lambda_3 \lambda}{\mt^4} + c_{30}\frac{\lambda_3^3}{\mt^6} + \left(c_{31} + \frac{9}{2\pi} c_{11} \right)\frac{\lambda_3^3 \lambda}{\mt^8} + \left(c_{50} + \frac{9}{2\pi} c_{30} \right)\frac{\lambda_3^5}{\mt^{10}} + \dots \,,
			\\
			\overline m^2(\mt,\lambda,\lambda_3)
			 & =  \mt^2 + 6 \, c_{11}\frac{\lambda_3^2 \lambda}{\mt^4} + 6 \, c_{30}\frac{\lambda_3^4}{\mt^6} + \dots \,.
		\end{aligned}
		\label{eq:param_expand}
	\end{equation}
	The expansion for $\overline\lambda_3$ is trivially obtained by the third eq.~in \eqref{eq:param_aux}.
	At this point we can obtain the perturbative series of our original theory $\Lagt$ by re-expanding the parameters $\overline m$ and $\overline\lambda_3$ in the series for the auxiliary theory $\overline {\cal L}$.  Moreover, since $\langle \varphi \rangle =0$, we have that the VEV is simply given by $\langle \phi \rangle = \phi_\mathrm{cl} + \phi_0\,$.
	From eq.~\eqref{eq:param_expand} we see that $\langle \phi \rangle$ is picking additional contributions with respect to the 1PI diagrams. At order $\lambda_3^3 \lambda$ and $\lambda_3^5$ they are
	\begin{align*}
		\raisebox{-.5\height}{\includegraphics[width=.7cm]{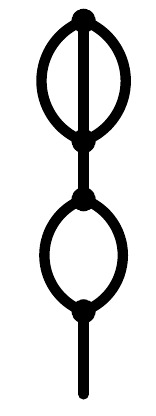}} \lambda_3^3 \lambda \,
		 & = \,
		\raisebox{-.5\height}{\rotatebox{0}{\includegraphics[height=.6cm]{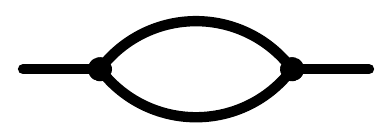}}} \, \lambda_3^2
		\cdot
		\raisebox{-.5\height}{\includegraphics[width=.9cm]{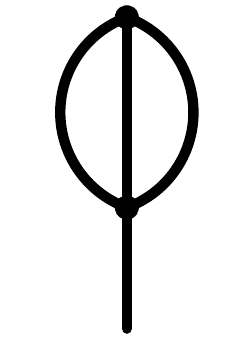}}\lambda_3 \lambda
		= \frac{9}{2\pi} \lambda_3^2 \cdot c_{11} \lambda_3 \lambda \,,
		\\
		\raisebox{-.5\height}{\includegraphics[width=.7cm]{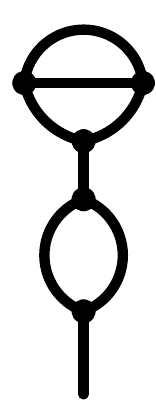}} \lambda_3^5 \,
		 & = \,
		\raisebox{-.5\height}{\rotatebox{0}{\includegraphics[height=.6cm]{figs/2pt-20.pdf}}} \, \lambda_3^2
		\cdot
		\raisebox{-.5\height}{\includegraphics[width=.9cm]{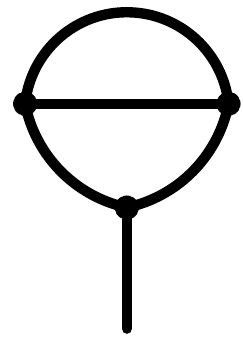}}\,\, \lambda_3^3
		= \frac{9}{2\pi} \lambda_3^2 \cdot c_{30} \lambda_3^3
		\,.
	\end{align*}
	Similarly, using the second eq.~in \eqref{eq:param_expand}, for the two point function in momentum space $\widetilde \Gamma_2 = p^2 +\overline m^2 +\dots$ we see that the expansion of $\overline m^2$ provides to lowest order the contribution of the following two graphs for the theory $\Lagt$
	\begin{equation*}
		\begin{split}
			\widetilde \Gamma_2 &=
			p^2 + \mt^2 + 6 \, c_{11}\frac{\lambda_3^2 \lambda}{\mt^4} + 6 \, c_{30}\frac{\lambda_3^4}{\mt^6} + \dots
			\\
			&= p^2 + \mt^2
			+ \raisebox{-.3\height}{\includegraphics[width=1.1cm]{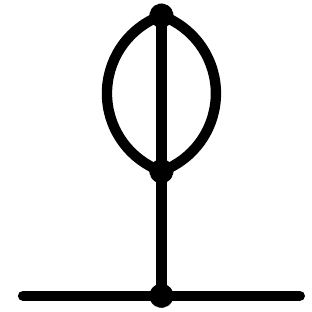}}\lambda_3^2 \lambda
			+ \raisebox{-.3\height}{\includegraphics[width=1.1cm]{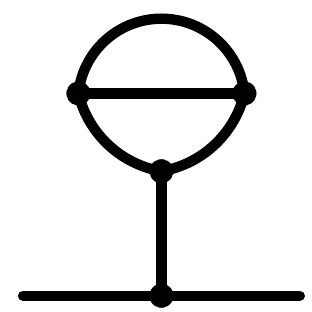}}\lambda_3^4 + \dots
			\,,
		\end{split}
	\end{equation*}
	which have to be added to the 1PI graphs at the same order (see eq.~\eqref{eq:G2_topo} below).
	Proceeding at higher order one must be careful in expanding also all the higher order terms of the expansion.

	In order to verify that the above procedure was correctly implemented, we checked that the series obtained for the VEV and for the 2-pt function matched a direct computation in the theory $\Lagt$ up to five loops.

	In the following we focus on the 0-, 1- and 2-point functions. Multi-loops computations have been addressed as in ref.~\cite{Serone:2018gjo} using the Montecarlo VEGAS algorithm \cite{Lepage:1977sw}.
	We refer the reader to ref.~\cite{Serone:2018gjo} for further details. We have computed Feynman diagrams up to the insertion of eight vertices, independently of the nature of the vertex, cubic or quartic.
	Since $\lambda_3 \sim \mt \sqrt{\lambda}$, this implies that our ordinary perturbative series can reach ${\cal O}(\gt^4)$.
	Using instead EPT as described in section~\ref{subsec:EPT}, where parametrically $\lambda_3^{{\rm EPT}} \sim \lambda$, we effectively can reach ${\cal O}(\gt^8)$.

	\subsection{Vacuum Energy}
	We have computed all the vacuum energy 1PI graphs with up to eight vertices in the auxiliary theory. The number of topologically distinct graphs (in the chosen scheme) as a function of the number of cubic and quartic diagrams is reported in tab.~\ref{tab:n-diag-0}.
	\begin{table}[t!]
		\centering
		\begin{tabular}{c|ccccccccc}
			\toprule
			$\Gamma_0$             & $\lambda^0$ & $\lambda^1$ & $\lambda^2$ & $\lambda^3$ & $\lambda^4$ & $\lambda^5$ & $\lambda^6$ & $\lambda^7$ & $\lambda^8$ \\
			\midrule
			$\overline\lambda_3^0$ & 0           & 0           & 1           & 1           & 3           & 6           & 19          & 50          & 204         \\
			$\overline\lambda_3^2$ & 1           & 1           & 4           & 12          & 54          & 232         & 1266                                    \\
			$\overline\lambda_3^4$ & 2           & 5           & 34          & 186         & 1318                                                                \\
			$\overline\lambda_3^6$ & 5           & 26          & 297                                                                                             \\
			$\overline\lambda_3^8$ & 16                                                                                                                          \\
			\bottomrule
		\end{tabular}
		\caption{Number of topologically distinct 1PI 0-pt diagrams without self-contractions with maximum eight total vertices for the auxiliary theory.}
		\label{tab:n-diag-0}
	\end{table}
	For illustration, the non-vanishing diagrams with up to three loops are
	\begin{equation}
		\Gamma_0 =
		- \,  \vcenter{\hbox{\includegraphics[width=1cm]{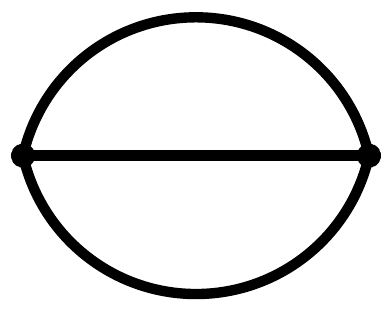}}} \, \overline\lambda_3^2
		- \left(
		\vcenter{\hbox{\includegraphics[width=1cm]{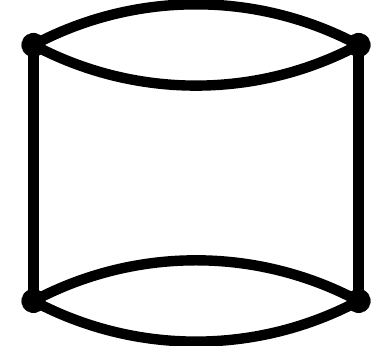}}}
		+  \vcenter{\hbox{\includegraphics[width=1cm]{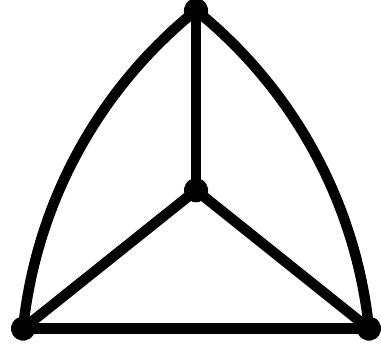}}} \,
		\right) \overline\lambda_3^4
		+  \vcenter{\hbox{\includegraphics[width=1cm]{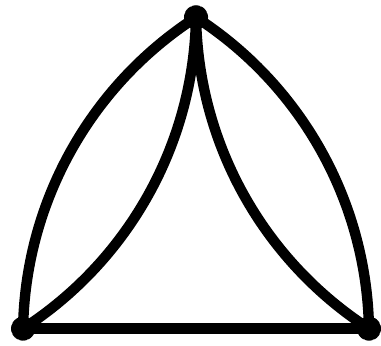}}} \, \overline\lambda_3^2 \lambda
		- \,  \vcenter{\hbox{\includegraphics[width=1.1cm]{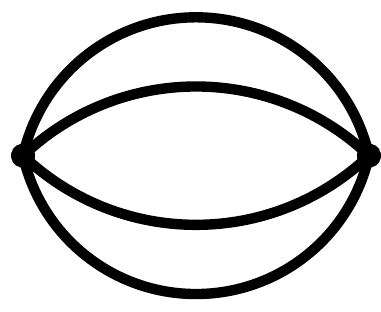}}} \, \lambda^2 + \dots \,.
	\end{equation}
	Expanding the parameters as explained above we get the following expression for the series of the vacuum energy
	\begin{equation}
		\frac{\Lt}{\widetilde m^2}
		=
		-\left(\frac{\psi^{(1)}(1/3)}{4\pi^2}  -\frac 16\right) \, \gt
		-0.042182971(51) \, \gt^2
		-0.0138715(74) \, \gt^3
		-0.01158(19) \, \gt^4
		+\mathcal{O}(\gt^5) \,,
		\label{eq:Lambda_series}
	\end{equation}
	where $\psi^{(n)}(z) = \partial_z^{(n+1)} \log \Gamma(z)$ is the polygamma function and the numbers in parenthesis indicate the error in the last two digits due to the numerical integration.
	The coefficients of the vacuum energy $\Lt$ for generic values of the couplings $\lambda_3$ and $\lambda$ are reported in the appendix (tab.~\ref{tab:Lambda-coeff}).
	In order to access the strong coupling regime of $\Lt$ we will use EPT as described in sec.~\ref{subsec:EPT}, therefore we report here the series obtained by setting $g_3 = \sqrt{2} \gt /\sqrt{\gt_0}$:
	\begin{equation}
		\begin{split}
			\frac{\Lt}{\widetilde m^2}
			\overset{\text{EPT}}{=}
			&- \left(\frac{1}{\gt_0}
			\left(\frac{\psi^{(1)}(1/3)}{4\pi^2}  -\frac 16\right)
			+\frac{21 \zeta (3)}{16 \pi ^3}\right) \gt^2
			+ \left(\frac{0.15991874}{\gt_0}
			+ \frac{27 \zeta (3)}{8 \pi ^4}\right) \gt^3\\
			&- \left(\frac{0.151218477(51)}{\gt_0^2}+\frac{0.75112786(68)}{\gt_0} +
			0.116125964(91)\right) \gt^4\\
			&+\left(\frac{1.8291267(28)}{\gt_0^2}+\frac{3.5560813(61)}{\gt_0}+0.3949534(18)\right)\gt^5\\
			&-\left(\frac{1.1335189(68)}{\gt_0^3}+\frac{16.41488(14)}{\gt_0^2}+\frac{18.827865(47)}{\gt_0} +1.629794(22) \right)\gt^6 \\
			&+ \left(\frac{24.4176(12)}{\gt_0^3}+\frac{138.643(10)}{\gt_0^2} +\frac{110.471(11)}{\gt_0} + 7.85404(21) \right)\gt^7\\
			&- \left(\frac{11.454254(57)}{\gt_0^4} + \frac{358.08(15)}{\gt_0^3} + \frac{1178.86(18)}{\gt_0^2} + \frac{712.76(72)}{\gt_0} + 43.192(21)  \right) \gt^8
			\,.
		\end{split}
		\label{eq:Lambda_EPT}
	\end{equation}

	\subsection{1-Point Tadpole}
	The series coefficients for the VEV have been obtained from $\widetilde \Gamma_1$ as explained above. We have computed all 1PI 1-pt graphs with up to eight vertices in the auxiliary theory. The number of topologically distinct graphs (in the chosen scheme) as a function of the number of cubic and quartic diagrams is reported in tab.~\ref{tab:n-diag-1}.
	\begin{table}[t!]
		\centering
		\begin{tabular}{c|cccccccc}
			\toprule
			$\Gamma_1$             & $\lambda^0$ & $\lambda^1$ & $\lambda^2$ & $\lambda^3$ & $\lambda^4$ & $\lambda^5$ & $\lambda^6$ & $\lambda^7$ \\
			\midrule
			$\overline\lambda_3^1$ & 0           & 1           & 2           & 6           & 23          & 95          & 464         & 2530        \\
			$\overline\lambda_3^3$ & 1           & 5           & 26          & 149         & 963         & 6653                                    \\
			$\overline\lambda_3^5$ & 3           & 29          & 302         & 2953                                                                \\
			$\overline\lambda_3^7$ & 12          & 223                                                                                             \\
			\bottomrule
		\end{tabular}
		\caption{Number of topologically distinct 1PI 1-pt diagrams without self-contractions with maximum eight total vertices for the auxiliary theory.}
		\label{tab:n-diag-1}
	\end{table}
	The non-vanishing diagrams with up to three loops have been already reported in eq.~\eqref{eq:G1_topo}.

	Following the method described at the beginning of sec.~\ref{sec:PertCoeff} we get the series for the vacuum expectation value of $\phi$ as
	\begin{equation}
		\frac{\langle \phi \rangle}{\phi_\mathrm{cl}}
		=
		1 - 0.712462426(83) \, \gt^2 - 2.152451(65)\, \gt^3 - 6.5422(59)\,\gt^4
		+ \mathcal{O}(\gt^5) \,,
		\label{eq:Tadpole_series}
	\end{equation}
	where $\phi_\mathrm{cl}$ is the tree-level value. The coefficients of $\langle \phi \rangle$ for generic values of the couplings $\lambda_3$ and $\lambda$ are reported in the appendix (tab.~\ref{tab:vev-coeff}).

	\subsection{Physical Mass}

	We define the physical mass as the smallest zero of the 1PI two-point function in momentum space
	for complex values of the Euclidean momentum:
	\be
	\widetilde \Gamma_2(p^2=-\Mt^2)\equiv 0\,.
	\label{mphDef}
	\ee
	By a perturbative expansion of $\Mt^2$ in powers of $\gt$, the zero of $\widetilde \Gamma_2(p^2=-\Mt^2)$ is determined in terms of
$\widetilde \Gamma_2(p^2=-\mt^2)$ and its derivatives with respect to $p^2$ (see ref.~\cite{Serone:2018gjo} for further details),
	which are in turn determined from $\widetilde \Gamma_2(p^2=-\overline m^2)$ and derivatives in the auxiliary theory. The number of topologically distinct 1PI graphs (in the chosen scheme) as a function of the number of cubic and quartic diagrams is reported in tab.~\ref{tab:n-diag-2}.
	\begin{table}[t!]
		\centering
		\begin{tabular}{c|ccccccccc}
			\toprule
			$\Gamma_2$             & $\lambda^0$ & $\lambda^1$ & $\lambda^2$ & $\lambda^3$ & $\lambda^4$ & $\lambda^5$ & $\lambda^6$ & $\lambda^7$ & $\lambda^8$ \\
			\midrule
			$\overline\lambda_3^0$ & 0           & 0           & 1           & 2           & 6           & 19          & 75          & 317         & 1622        \\
			$\overline\lambda_3^2$ & 1           & 3           & 14          & 61          & 342         & 2018        & 13499                                   \\
			$\overline\lambda_3^4$ & 2           & 17          & 163         & 1400        & 12768                                                               \\
			$\overline\lambda_3^6$ & 9           & 136         & 2177                                                                                            \\
			$\overline\lambda_3^8$ & 46                                                                                                                          \\
			\bottomrule
		\end{tabular}
		\caption{Number of topologically distinct 1PI 2-pt diagrams without self-contractions with maximum eight total vertices for the auxiliary theory.}
		\label{tab:n-diag-2}
	\end{table}
	For illustration, the non-vanishing diagrams with up to two loops are
	\begin{equation}
		\begin{split}
			\widetilde \Gamma_2 &=
			p^2 + \overline m^2
			- \vcenter{\hbox{\includegraphics[width=1cm]{figs/2pt-20}}} \, \overline\lambda_3^2
			- \left[
				\vcenter{\hbox{\includegraphics[width=1cm]{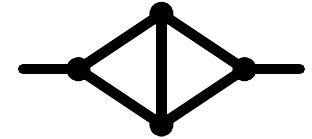}}}
				+ \vcenter{\hbox{\includegraphics[width=1cm]{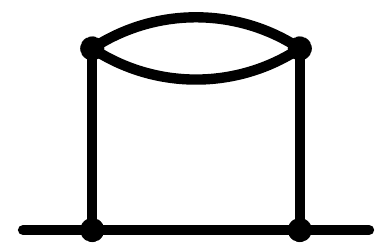}}} \,
				\right] \overline\lambda_3^4
			- \vcenter{\hbox{\includegraphics[width=1.1cm]{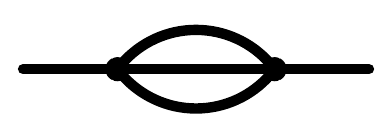}}} \, \lambda^2
			\\
			& + \left[
				\vcenter{\hbox{\includegraphics[width=1cm]{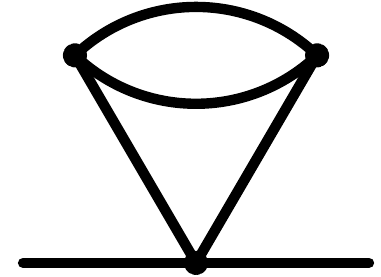}}} +
				\vcenter{\hbox{\includegraphics[width=1cm]{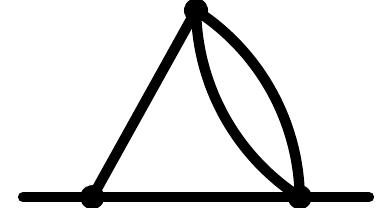}}} +
				\vcenter{\hbox{\includegraphics[width=1.5cm]{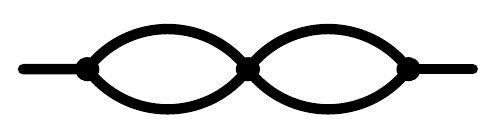}}}
				\right] \, \overline\lambda_3^2 \lambda
			+\dots
			\,.
		\end{split}
		\label{eq:G2_topo}
	\end{equation}
	In tab.~\ref{tab:Gamma2n-coeff} in the appendix we report the coefficients for $\widetilde \Gamma_2(p^2=-\mt^2)$ and its derivatives for generic values of the couplings $\lambda_3$ and $\lambda$. For $\lambda_3 = \sqrt{2 \lambda}\mt$ we get the following expression for the physical mass
	\begin{equation}
		\frac{\Mt^2}{\mt^2}
		= 1 -2 \sqrt{3} \, \gt -4.1529(18) \gt^2 -14.886(30) \gt^3 - 50.62(99) \gt^4 +\mathcal{O}(\gt^5) \,.
		\label{eq:Mass_series}
	\end{equation}

	\subsection{Large Order Behavior}

	The large order behavior of the perturbative expansion of $n$-point Schwinger functions $G_n$ in the 2d $\phi^4$ theory in the broken phase has not been determined before.
	On general grounds, we expect that the coupling expansion $G_n = \sum_k G_n^{(k)} \gt^k$  behaves, for $k\gg 1$, as
	\be
	G_n^{(k)}  = c_n (a+a^*)^k \Gamma(k+ b_n+1)\Big(1+{\cal O}(k^{-1}) \Big) \,,
	\label{LOBBF}
	\ee
	where $c_n$ and $b_n$ are $n$-dependent constants, while $a$ is an $n$-independent constant given
	by the inverse of $t_1^-$ in eq.~(\ref{eq:tipm}):
	\be
	a \approx 0.423096 + 0.450462 \,i \,, \quad a\equiv \frac{1}{t_1^-}\,.
	\ee
	The half-period of oscillation of the large order coefficients is given by $\pi/arg(a)\simeq3.8$.
	The evaluation of the coefficients $b_n$ and $c_n$ require a detailed analysis of small fluctuations around the instanton configuration which we have not attempted to perform.
	However, the knowledge of $a$ is enough to allow us to use a conformal mapping, as discussed in sec.~\ref{subsec:NCM}.

	The shortness of our perturbative series does not allow for a reliable comparison with the asymptotic large order behavior. We just observe that the period of oscillation
	given by eq.~(\ref{LOBBF}) is consistent  with the same-sign behavior of the first terms in eqs.~(\ref{eq:Lambda_series}), (\ref{eq:Tadpole_series}) and (\ref{eq:Mass_series}).

	\section{Results}

	\label{sec:results}

	We report in this section the numerical results obtained from the truncated perturbative series and from its Borel resummation.
	We have considered conformal mapping and Pad\'e-Borel approximants methods to obtain a numerical estimate of the Borel function.
	Most of the details of our numerical implementation, as well as a short introduction to these resummation methods, has been given in ref.~\cite{Serone:2018gjo},
	so here we will only focus on some characteristic features of the broken phase.

	The generalized conformal mapping method explained in section \ref{subsec:NCM} allows us to write any observable $F(\gt)$ using eqs.~(\ref{eq:borelmap}) and (\ref{eq:compmap}),
	with $|t_1^\pm|$ and $\alpha_1$ the modulus and phase of the inverse of the leading (complex) instanton action in eq.~(\ref{eq:tipm}). In terms of the Le Roy - Borel function
	\be
	{\cal B}_b(t) = \sum_{n=0}^\infty \frac{F_n}{\Gamma(n+b+1)} t^n \equiv \sum_{n=0}^\infty B_n^{(b)} t^n  \,,
	\label{LeRoy}
	\ee
	we have
	\bea
	F_B(\gt) & = & \frac 1{\gt} \int_0^\infty \!dt \, \Big(\frac{t}{\gt}\Big)^b e^{-t/\gt} \sum_{n=0}^\infty B_n^{(b)} t^n =
	\frac{1}{\gt^{b+1}}\int_0^1 \!du \, \frac{dt}{du} e^{-t(u)/\gt} \frac{t^b(u)}{(1-u)^{2s}} \sum_{n=0}^\infty \widetilde B_n^{(b,s)} u^n \nn \\
	& \sim  & \frac{1}{\gt^{b+1}}\sum_{n=0}^N \widetilde B_n^{(b,s)}  \int_0^1 \!du \, \frac{dt}{du} e^{-t(u)/\gt} \frac{t^b(u) u^n}{(1-u)^{2s}}  \,.
	\label{CMtTou}
	\eea
	In order to not clutter the notation, we omit to write the $(b,s)$ dependence  of $F_B(\gt)$.
	As in ref.~\cite{Serone:2018gjo}, we introduced two summation variables denoted by $b$ and $s$  to further improve the behavior of the $u$-series and to have more control on the accuracy of the results.
	Due to the presence of additional singularities within the unit $u$-disc, after the conformal mapping (\ref{eq:compmap}) the series is still asymptotic. The conformal mapping in this case is supposed to extend the region in coupling space where the asymptotic series behaves effectively as a convergent one, giving a more reliable estimate of the observable with respect to optimal truncation of the perturbative series. It is useful to estimate the rate of ``convergence" of the series after conformal mapping to have a rough expectation on the accuracy of the results
	obtained with finite truncations. In order to simplify our discussion, we might consider the ideal situation where no additional singularities are present. In this case the series in the last line of eq.~(\ref{CMtTou}) is truly convergent for any value of the coupling constant and its rate of convergence is governed by the large $n$ behavior of the $u$ integral. For $n\gg 1$ a saddle point approximation gives
	\be
	\int_0^1 \!du \, \frac{dt}{du} e^{-t(u)/\gt} \frac{t^b(u) u^n}{(1-u)^{2s}} \approx C \exp\Big(-c \frac{n^{\frac{2\alpha_1}{2\alpha_1+1}}}{(|a| \gt)^{\frac{1}{2\alpha_1+1}}}\Big)\,,
	\label{CMestimate}
	\ee
	where $\alpha_1$ is the angle between the real positive axis and the leading singularity in the Borel $t$-plane appearing in eq.~(\ref{eq:compmap}),
$C$ is a $b$-, $s$- and $\alpha_1$-dependent coefficient which is polynomial in $n$, and $c$ is a smooth function of $\alpha_1$ of order one for all values of $|\alpha_1|\leq 1$. We see that the exponential convergence in $n$ of the $u$ integral sensitively depends on $\alpha_1$. It is maximal for $\alpha_1 = 1$, it
	monotonically decreases for smaller values of $\alpha_1$ and eventually vanishes for $\alpha_1=0$, when the series is no longer Borel resummable. In the $\mathbb Z_2$ unbroken and broken theories we have $\alpha_1 = 1$, $\alpha_1 \simeq 0.260$ respectively.  At fixed number of orders we then expect a slower convergence of the conformal mapping method in the broken case with respect to the unbroken one.

	Of course the overall convergence of the series in $n$ in eq.~(\ref{CMtTou}) is also determined by the large order behavior of the coefficients $\widetilde B_{n}^{(b,s)}$, which is governed by the other singularities in the Borel plane. In the unbroken theory the known singularities are all aligned with the leading one along the negative real axis, the best case scenario, while we have explicitly seen in section \ref{subsec:NCM} that this is not the case for the broken case.
	We then expect that next to leading singularities would most likely further increase the gap in accuracy between resummations in the unbroken and broken phases, though
	we believe that for $\gt \leq \gt_c^{(w)}$ this is a sub-dominant effect.

	Due to the shortness of the series we obtained in sec.~\ref{sec:PertCoeff}, the error minimization procedure of ref.~\cite{Serone:2018gjo} that is used to select the central values for the parameters $b$ and $s$
	is sometimes unstable (especially for the 2-point function). We also note that in the broken phase the error minimization procedure tends to select lower values of $b$ than in the unbroken phase. In order to avoid the dangerous $b=-1$ point at which the $\Gamma$-function in the Borel-Le Roy transform diverges, in this paper we fix $\Delta b=1$, where $\Delta b$ is the semirange used to scan the parameter $b$ for the error estimation as in ref.~\cite{Serone:2018gjo}.
	The resulting error estimate typically yields large errorbars but this is expected since we are only resumming few perturbative terms. Tests on simple toy models showed that the error estimate usually correctly represents the difference between the resummed and true values.

	In contrast to the unbroken phase, Pad\'e-Borel approximants are not very useful because the same-sign form of the first available series coefficients
	is typically responsible for spurious unphysical poles that hinder a proper use of this method. Moreover, the shortness of the series does not allow us to
	systematically select the ``best" Pad\'e-Borel approximant as explained in ref.~\cite{Serone:2018gjo}. Nevertheless, we report for completeness the
	results obtained using Pad\'e-Borel approximants, when available.
	The shortness of the series does not allow us to estimate the contribution to the error coming from the convergence.

	We have also computed the vacuum energy using EPT as explained in section \ref{subsec:EPT}. In this case the leading singularity of the Borel function  ${\cal B}(t,\gt_0)$, at fixed $\gt_0$, is the same as in the unbroken phase, with $\alpha_1=1$, $|a| \simeq 0.683708$. We now have
	\bea
	\hat F_B(\gt,\gt_0) & = & \frac 1{\gt} \int_0^\infty \!dt \, \Big(\frac{t}{\gt}\Big)^b e^{-t/\gt} \sum_{n=0}^\infty B_n^{(b,\gt_0)} t^n =
	\frac{1}{\gt^{b+1}}\int_0^1 \!du \, \frac{dt}{du} e^{-t(u)/\gt} \frac{t^b(u)}{(1-u)^{2s}} \sum_{n=0}^\infty \widetilde B_n^{(b,s)}(\gt_0) u^n \nn \\
	& \sim  & \frac{1}{\gt^{b+1}}\sum_{n=0}^N \widetilde B_n^{(b,s)}(\gt_0)  \int_0^1 \!du \, \frac{dt}{du} e^{-t(u)/\gt} \frac{t^b(u) u^n}{(1-u)^{2s}}  \,.
	\label{CMtTouEPT}
	\eea
	The value of an observable $F_B(\gt)$ is recovered by eventually setting $\gt_0 = \gt$ in eq.~(\ref{CMtTouEPT}):
	\be
	\hat F_B(\gt,\gt) = F_B(\gt).
	\ee
	The different analytic structure of the Borel function and a longer perturbative series allows us to use Pad\'e-Borel approximants in EPT.
	Unless stated differently, we set in what follows $\widetilde m^2=1$.

	\subsection{Vacuum Energy: Weak Coupling}

	\begin{figure}[t!]
		\centering
		\includegraphics[width=.48\textwidth]{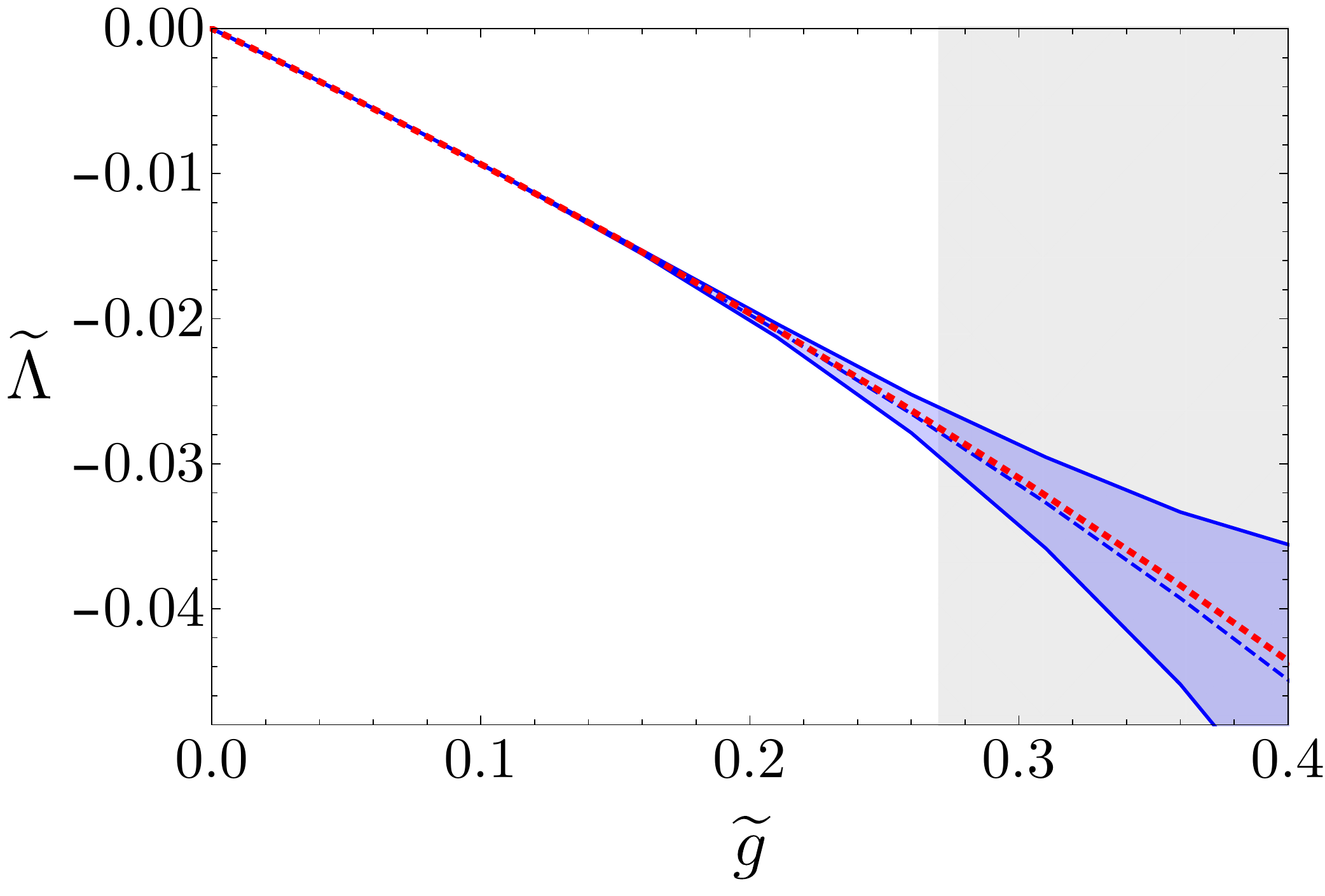}~~
		\includegraphics[width=.48\textwidth]{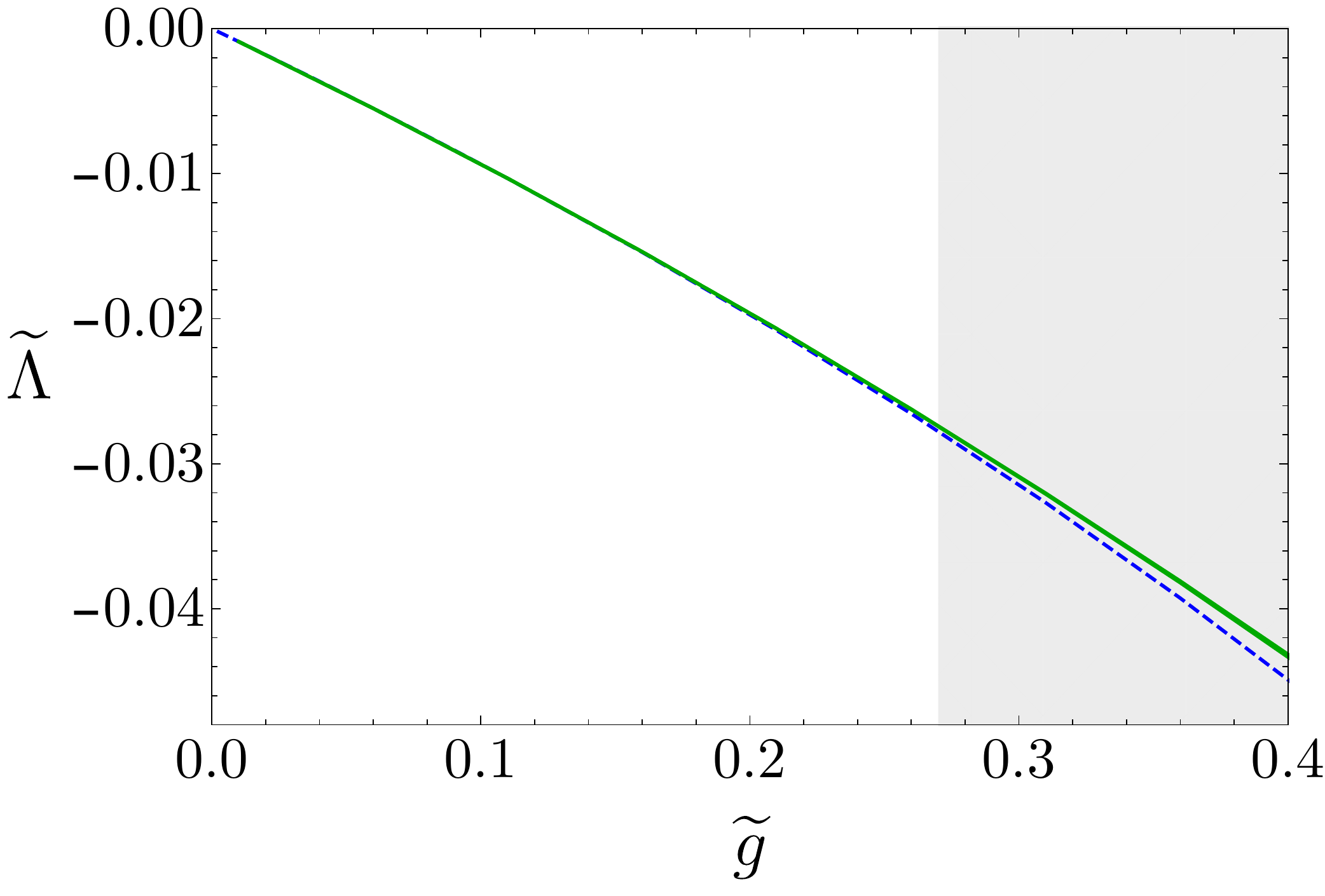}
		\caption{The vacuum energy $\Lt$ as a function of the coupling constant $\gt$ using the coefficients up to the $\gt^4$ order. Left panel: the results obtained using the conformal mapping \eqref{CMtTou} (light blue) with parameters $s=1/5$, $b=0$ and the optimally truncated series (red dotted line)  which coincides with the perturbative series up to $\gt^4$ for the couplings shown in the plot. Right panel: comparison between the central values of the conformal mapping (blue dashed line) and Padé-Borel (solid green) resummation techniques. The regions beyond the phase transition $\gt> \gt_c^{(w)}$ are shaded in gray.}
		\label{fig:LambdaPT}
	\end{figure}
	The perturbative expression for $\Lt$ up to order $\gt^4$ is reported in eq.~(\ref{eq:Lambda_series}).
	We show in the left panel of fig.~\ref{fig:LambdaPT} $\Lt(\gt)$ as a function of $\gt$. Surprisingly enough, the vacuum energy series is within the perturbative regime
	up to the critical coupling $\gt_c$ and well beyond, as evident from the fact that optimal truncation and untruncated perturbation theory coincides for all the couplings shown in the figure and in this whole regime they well approximate the central value of the Borel resummed result (blue dashed line). For $\gt\gtrsim 0.2$, the error associated to the resummation rapidly increases, despite the central values remain quite close.

	In the right panel of fig.~\ref{fig:LambdaPT} we compare $\Lt (\gt)$ computed using conformal mapping and Padé-Borel resummation techniques.
	The results with the conformal mapping \eqref{CMtTou} are obtained using resummation parameters $s=1/5$ and $b=0$. The approximant used
	in the Pad\`e-Borel method is $[1/2]$ with parameter $b=1$.
	The results are all well compatible  with each other, confirming that the vacuum energy is perturbatively accessible in the whole
	range of the weakly coupled branch.

	\subsection{Vacuum Energy: Strong Coupling and Chang Duality Checks}

\begin{figure}[t!]
		\centering
		\includegraphics[width=.6\textwidth]{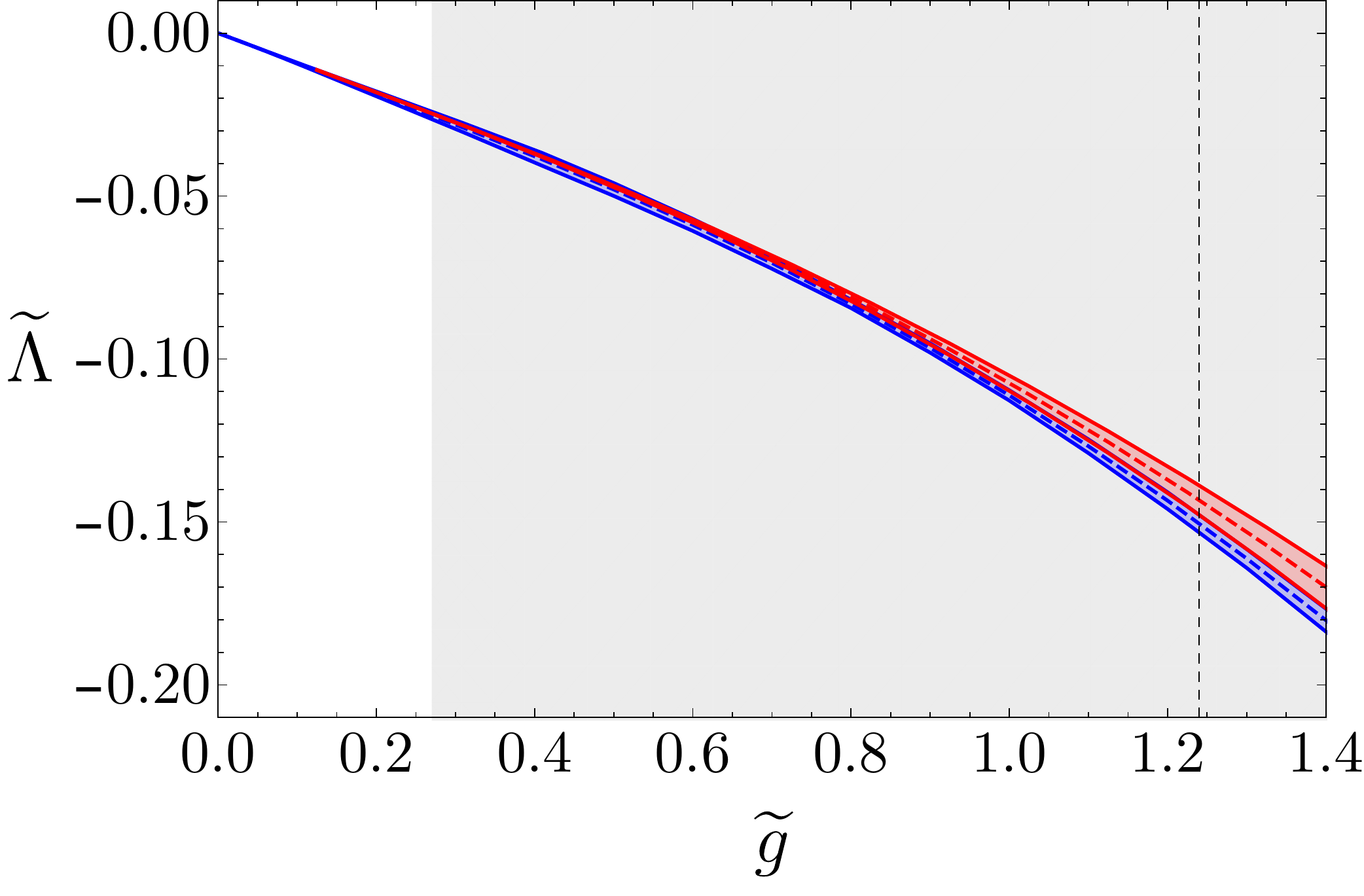}
		\caption{The vacuum energy $\Lt$ at strong coupling obtained by resummation of the EPT series~\eqref{eq:Lambda_EPT} with the conformal mapping (light blue) and Padé-Borel (light red) methods. The region beyond the phase transition $\gt> \gt_c^{(w)}$ is shaded in gray and the vertical dashed line signals the critical coupling in the strong branch $\gt = \gt_c^{(s)}$. }
		\label{fig:Lambda_EPT}
	\end{figure}

	The perturbative expression for $\Lt$ up to order $\gt^8$ in EPT is reported in eq.~(\ref{eq:Lambda_EPT}).
	As discussed in detail in ref.~\cite{Serone:2017nmd}, EPT works at its best at strong coupling. More quantitatively, at fixed number of loops $N$, we expect that EPT
	improves over ordinary perturbation theory when $g^2 > 1/N$.\footnote{This estimate has been established for ordinary integrals and numerically checked in quartic oscillators in quantum mechanics. We assume here
		that it qualitatively also holds in the 2d case.} It is then the ideal tool to compute the vacuum energy at strong coupling (strong branch in the broken phase). We report in fig.~\ref{fig:Lambda_EPT} $\Lt$ as a function of $\gt$  and compare the results obtained using conformal mapping and Pad\'e-Borel approximants. The light blue line corresponds to the conformal mapping at $N=8$. In order to avoid dangerous poles, in the Pad\'e-Borel method (light red) we have removed the vanishing ${\cal O}(\gt^0)$ and ${\cal O}(\gt)$ coefficients from the series and effectively resummed $\Lt(\gt)/\gt^2$. The approximant shown is $[3/2]$ with $b=-1/2$.
	The results are in good agreement with each other.

\begin{figure}[t!]
		\centering
		\includegraphics[width=.65\textwidth]{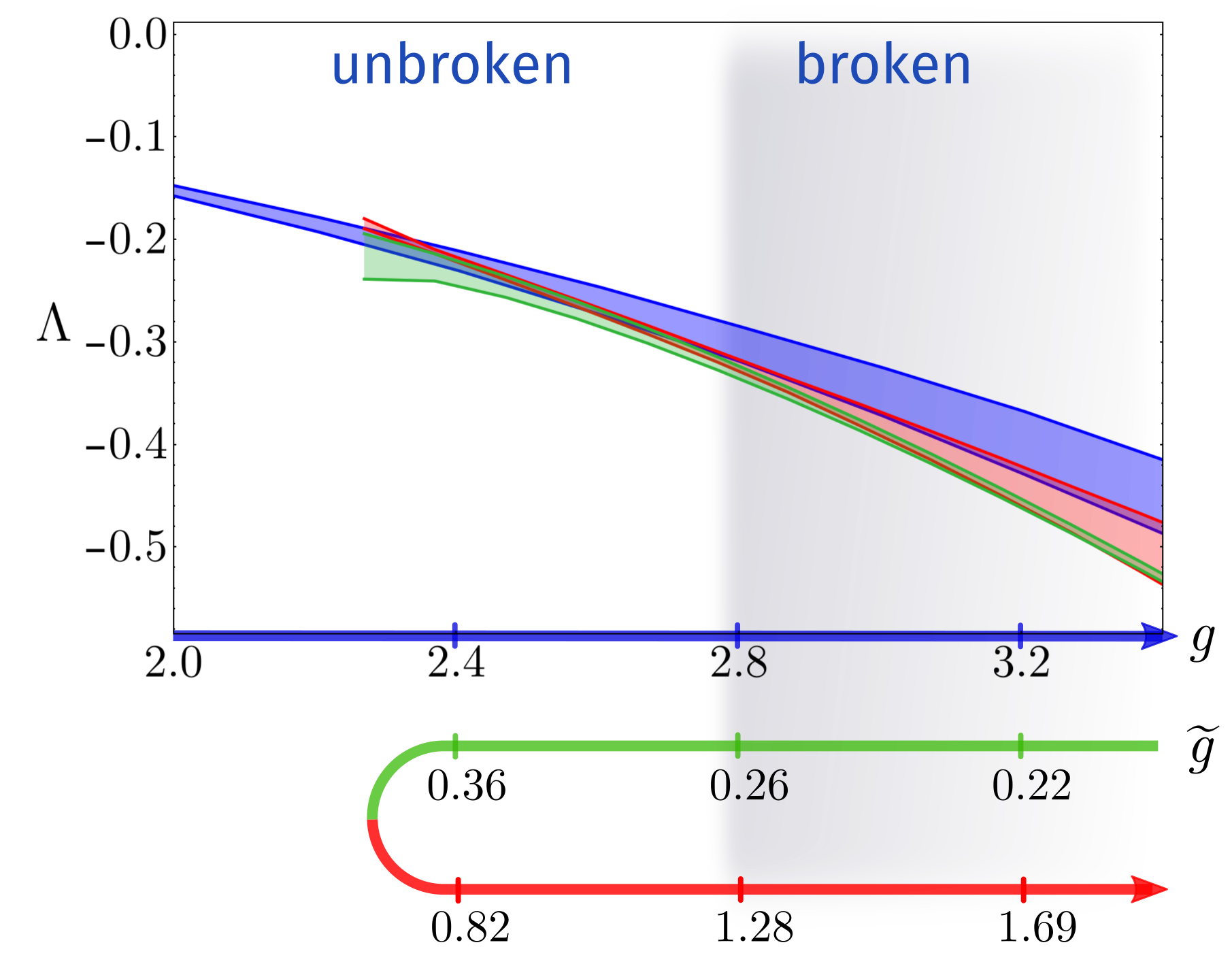}
		\caption{The vacuum energy $\Lambda$ for the theory described by the Lagrangian~\eqref{eq:LagI} as a function of the coupling constants $g$ (blue), $\gt^{(w)}$ (green) and $\gt^{(s)}$ (red).
			The blue band is computed by resumming the series in the unbroken phase (conformal mapping as in ref.~\cite{Serone:2018gjo}). The green band is obtained via Chang duality from the weak branch of the broken phase (conformal mapping in eq.~(\ref{eq:compmap}))). The red band is obtained via Chang duality from the strong branch using the EPT series in eq.~\eqref{eq:Lambda_EPT}  (conformal mapping as in ref.~\cite{Serone:2018gjo}).}
		\label{fig:LambdaChang}
	\end{figure}

	We can numerically check Chang duality in its full glory by comparing the vacuum energies in the unbroken, and weak/strong branches of the broken phases for values of the couplings
	associated to the same physical theory (points in the same vertical line in	fig.~\ref{fig:ChangPic}).
	Indeed, it has been argued in ref.~\cite{Serone:2018gjo} that the Borel resummation of perturbation theory around the unbroken vacuum,
	when applied beyond the phase transition point and without a proper selection of the vacuum,
	reconstructs the correlation functions $F$  in a vacuum where cluster decomposition is violated.\footnote{Note that the presence of a branch point \cite{Onsager:1943jn} for $\Lambda$ at $g=g_c$ might spoil the analytic continuation of the Borel resummation beyond the phase transition.
		Our results show no sign of such a failure, although the weakness of the non-analyticity
		might require a higher level of precision to become manifest.}
	However, since the vacua $|\pm\rangle$ are degenerate, the vacuum energy $\Lambda$ computed for $g\geq g_c$
	starting from the unbroken phase, coincides with the vacuum energies as computed from the broken phase in the weak and the strong branches.
	We summarize our findings in fig.~\ref{fig:LambdaChang}.
	We report the vacuum energy as a function of the coupling constants in the various phases, $g$ in the unbroken phase (blue), $\gt^{(w)}$ in the weakly branch of the broken phase (green) and $\gt^{(s)}$ in the strong branch of the broken phase (red). In comparing the vacuum energy in different phases one has to pay attention to the different units of mass and
	vacuum energy normalizations in the three descriptions, which are related as in eq.~(\ref{eq:Chang}). In fig.~\ref{fig:LambdaChang} we have set to unity the
	squared mass term $m^2$ (so that $\widetilde m^2 \neq 1$ in both  weak and strong branches) and normalized the vacuum energy $\Lambda$ to be zero for $g=0$ in the unbroken phase.
	The three vacuum energies are consistent with each other, as expected from Chang duality. We consider this result a numerical check of the Borel summability of the
$\phi^4$ theory in the broken phase and an example of the use of EPT in QFT.

	\subsection{Tadpole}

	\label{subsec:tadpole}

	The perturbative expression for $\langle \phi \rangle$, normalized to its classical value, up to order $\gt^4$ is reported in eq.~(\ref{eq:Tadpole_series}).
	We have resummed the expression $\langle \phi \rangle/\phi_\mathrm{cl}$ to the eighth power, because it shows better convergence properties than $\langle \phi \rangle/\phi_\mathrm{cl}$.
	This is not surprising. We know that in the 2d Ising model $\Mt \propto |\gt_c - \gt|$ (critical exponent $\nu =1$) and $\langle \phi \rangle \propto \Mt^{1/8}$ (critical exponent $\beta=1/8$), so that  $T\equiv (\langle \phi \rangle/\phi_\mathrm{cl})^8\propto  |\gt_c - \gt|$ approaches the critical coupling in an analytic way.\footnote{Of course, we are relying here on the knowledge of the critical exponents of the Ising model as input.}

	\begin{figure}[t!]
		\centering
		\includegraphics[width=.6\textwidth]{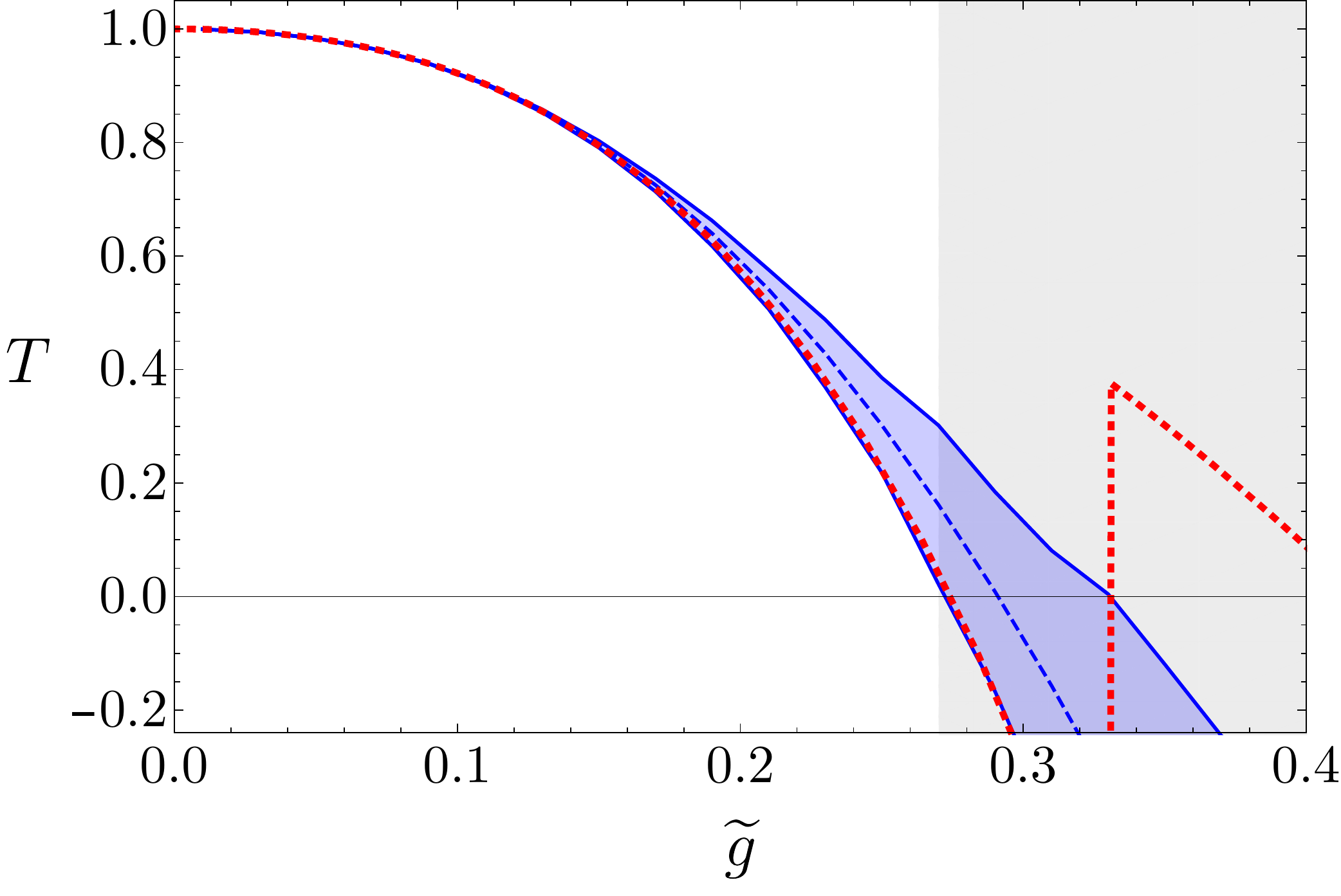}
		\caption{The quantity $T = (\langle \phi \rangle/ \phi_\mathrm{cl})^8$ as a function of the coupling constant $\gt$ in the weakly coupled branch using the coefficients up to the $\gt^4$ order. We compare the Borel resummation result using conformal mapping (blue dashed line) with optimal truncation (red dotted line).
			Note how the perturbative series gives a reliable result all the way up to $\gt_c^{(w)}$.}
		\label{fig:tad}
	\end{figure}

	We show in fig.~\ref{fig:tad} $T$ as a function of $\gt$ in the weak coupling regime. The series for $T$ is within the perturbative regime up to the critical coupling $\gt_c$ and soon after breaks down.
	The value of $\gt_c^{(w)}$ perturbatively found using the 4-loop series is surprisingly closer to the value expected from Chang duality $\gt_c^{(w)}\approx 0.27$. It would be nice to compute
	the series up to 5-loops to establish if this is a mere coincidence or not. The blue dashed line represents the central value of $T$ obtained using the conformal mapping with resummation parameters $s=1/4$ and $b=3/2$. The results are consistent with perturbation theory, but the resummation allows us to better estimate the error. Using the results of our resummation we get
	\be
	\gt_c^{(w)} = 0.29\pm 0.02 \,, \quad \quad \text{(from tadpole)}
	\label{eq:gcfromT}
	\ee
	in good agreement with the value (\ref{eq:gcfromChang}) derived using Chang duality from $g_c$ computed from the unbroken phase.
	The values of $T$ for $\gt \gtrsim \gt_c^{(w)}$ do not have an immediate physical meaning.
	At $\gt_c^{(w)}$ the vacua $|\pm\rangle$ collide and result in the single $\mathbb{Z}_2$ invariant vacuum where $T=0$ identically.

	\subsection{Mass}
	\label{subsec:massresults}

\begin{figure}[t!]
		\centering
		\includegraphics[width=.6\textwidth]{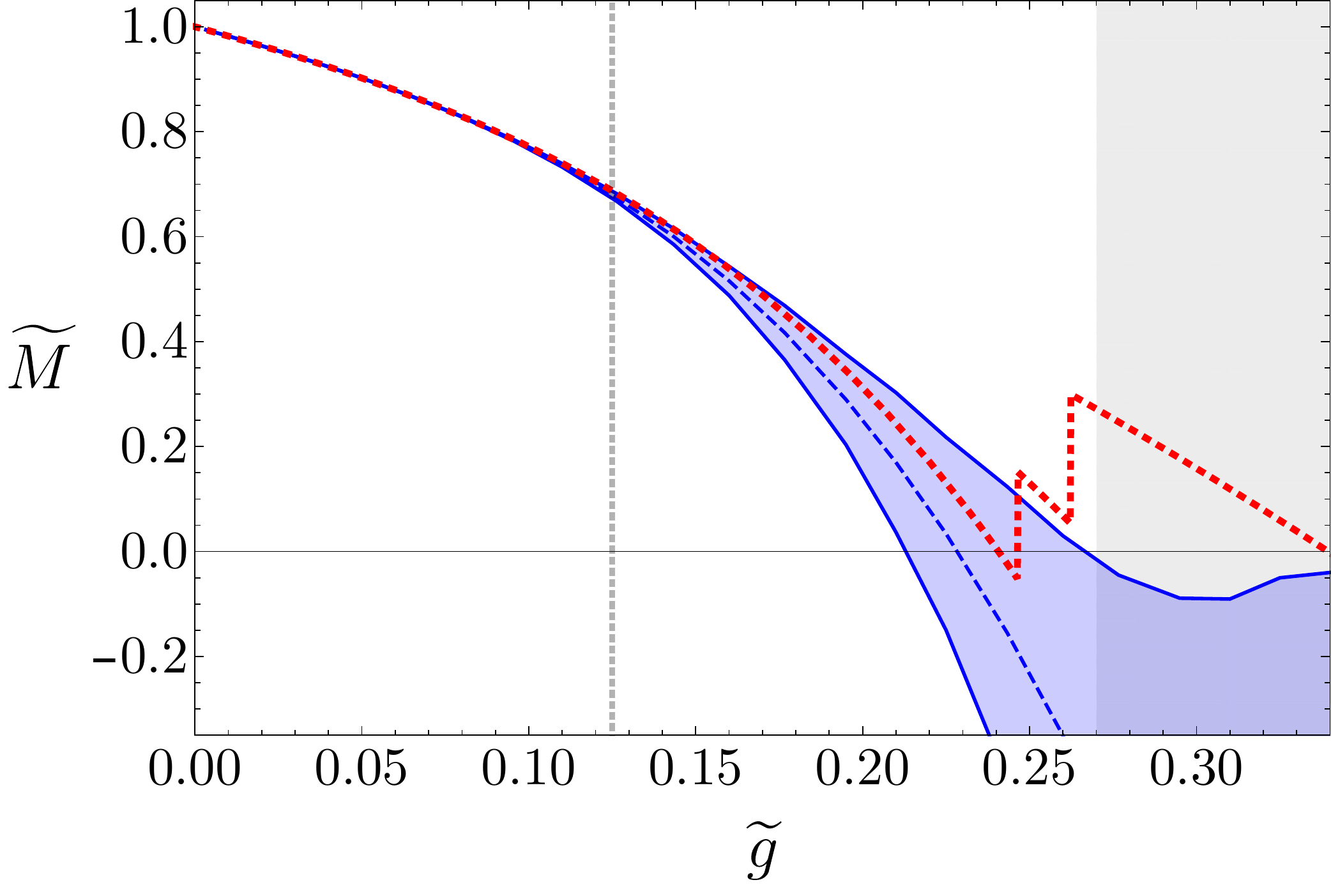}
		\caption{The physical mass $\Mt$ as a function of the coupling constant $\gt$ in the weakly coupled branch using the coefficients up to the $\gt^4$ order.
			We compare the Borel resummation result using conformal mapping (blue dashed line) with optimal truncation (red dotted line). The vertical dotted line at $\gt = \gt_{k \bar k}\approx 0.125$ is the coupling (taken from ref.~\cite{Bajnok:2015bgw}) where the mass of the elementary particle is  twice the mass of the kink and might become unstable.
		}
		\label{fig:mass}
	\end{figure}

	The perturbative expression for $\Mt $ up to order $\gt^4$ is reported in eq.~(\ref{eq:Mass_series}).
	We show in fig.~\ref{fig:mass} $\Mt$ as a function of $\gt$ in the weak coupling regime.
	The series for $\Mt$ is within the perturbative regime up to $\gt \gtrsim 0.2$ but, in contrast to $\widetilde \Lambda$ and $T$, it breaks down slightly before reaching $\gt_c^{(w)}$.
	The blue dotted line represents the central value of $\Mt$ obtained using conformal mapping
	with resummation parameters $s=1/2$ and $b=2$.
	The interpretation of $\Mt$ beyond a certain value of the coupling $\gt_{k\bar k}<\gt_c^{(w)}$ is tricky and is postponed to next section where a comparison with hamiltonian truncation methods is made.
	Independently of its physical interpretation, note however that $\Mt$ vanishes for a value of the coupling of roughly $0.23$, close to the value of  $\gt_c^{(w)}$ obtained from $T$ in eq.~(\ref{eq:gcfromT}) and from the one in eq.~(\ref{eq:gcfromChang}) expected from Chang duality.

	\section{Comparison with Refs.\cite{Rychkov:2015vap,Bajnok:2015bgw} and Mass Interpretation}
	\label{sec:comparisons}

	Before comparing our results with those of refs.\cite{Rychkov:2015vap,Bajnok:2015bgw}, it is useful to briefly recall basic facts about the Hilbert space structure of the $\phi^4$ theory in the broken phase.
	The Hamiltonian truncation methods of refs.\cite{Rychkov:2015vap,Bajnok:2015bgw} are based on the study of the spectrum of the $\phi^4$ theory defined on a spatial circle $S^1$ of circumference $L$.
	On $\mathbb R \times S^1$ the Hilbert space of the theory  is divided in two subsectors, according to the periodicity conditions of $\phi$,
	periodic or antiperiodic, around $S^1$. The lowest energy state in the Hilbert space corresponds to the ground state of the periodic sector, while the ground state in the antiperiodic sector
	is identified with the kink state in the infinite length limit $L\rightarrow\infty$. No spontaneous symmetry breaking can occur at finite $L$, since the vacuum is a unique state linear combination of $ |+\rangle$ and $ |-\rangle$, where $|\pm\rangle$ denote the two vacua where at tree-level $\langle \phi \rangle = \pm v$. The periodic Hilbert space sector is characterized by a quasi degenerate spectrum of states, whose energy splitting
	decreases exponentially with $L$ and is governed by the energy of the antiperiodic vacuum, i.e. the kink mass. In particular, for $L$ large enough, the lowest energy state beyond the vacuum is the combination of $|+\rangle$ and $|-\rangle$
	orthogonal to the vacuum, which becomes degenerate with it for  $L\rightarrow\infty$.
	In this limit  all states in the periodic sector become exactly degenerate and the Hilbert space is expected to also contain states that can be seen as composed of an even number of kink and anti-kink states.\footnote{Indeed a kink and an anti-kink, when far apart, are approximate finite energy solutions to the classical equations of motion, so it is natural to expect state configurations of this kind.} Analogously, states in the antiperiodic sector can be interpreted as composed of an odd number of kink and anti-kink states.
	A superselection rule forbids transitions that do not preserve a $\mathbb Z_2$ topological charge, the kink number operator.
	Semi-classical arguments \cite{Mussardo:2006iv} suggest that the elementary $\phi$-particle excitation decays into a pair of kink anti-kink states at some value of the coupling $\gt_{k\bar k}<\gt_c^{(w)}$.
	The presence of such decay has been checked numerically in ref.~\cite{Bajnok:2015bgw} (see also ref.~\cite{Coser:2014lla}), where the absence of single particle states in the periodic sector for $\gt_{k\bar k}\approx 0.125$
	is interpreted as its decay in a pair of kink and anti-kink states.

	The $\phi^4$ theory discussed in refs.\cite{Rychkov:2015vap,Bajnok:2015bgw}, in the $L\rightarrow\infty$ limit, becomes  a non-compact theory with two degenerate vacua connected by
	topological kink excitations. No vacuum selection has been performed, and therefore $\langle \phi \rangle = 0$ and no spontaneous symmetry breaking can occur.
	In contrast, our results are based on ordinary perturbation theory in non-compact space, expanding around $|+\rangle$ or $|-\rangle$, where spontaneous symmetry breaking occurs.
	This point should be taken into account when comparing our results with those of refs.\cite{Rychkov:2015vap,Bajnok:2015bgw}.
	We do not expect subtleties related to the choice of vacuum for the vacuum energy $\widetilde \Lambda$, since this is a continuous and smooth function at least up to $\gt_c^{(w)}$.
	Similarly, $\Mt$ is a smooth function as long as the particle is stable and should coincide with the mass of the lightest single particle state for $\gt <\gt_{k\bar k}$.

	\begin{figure}[t!]
		\centering
		\includegraphics[width=.47\textwidth]{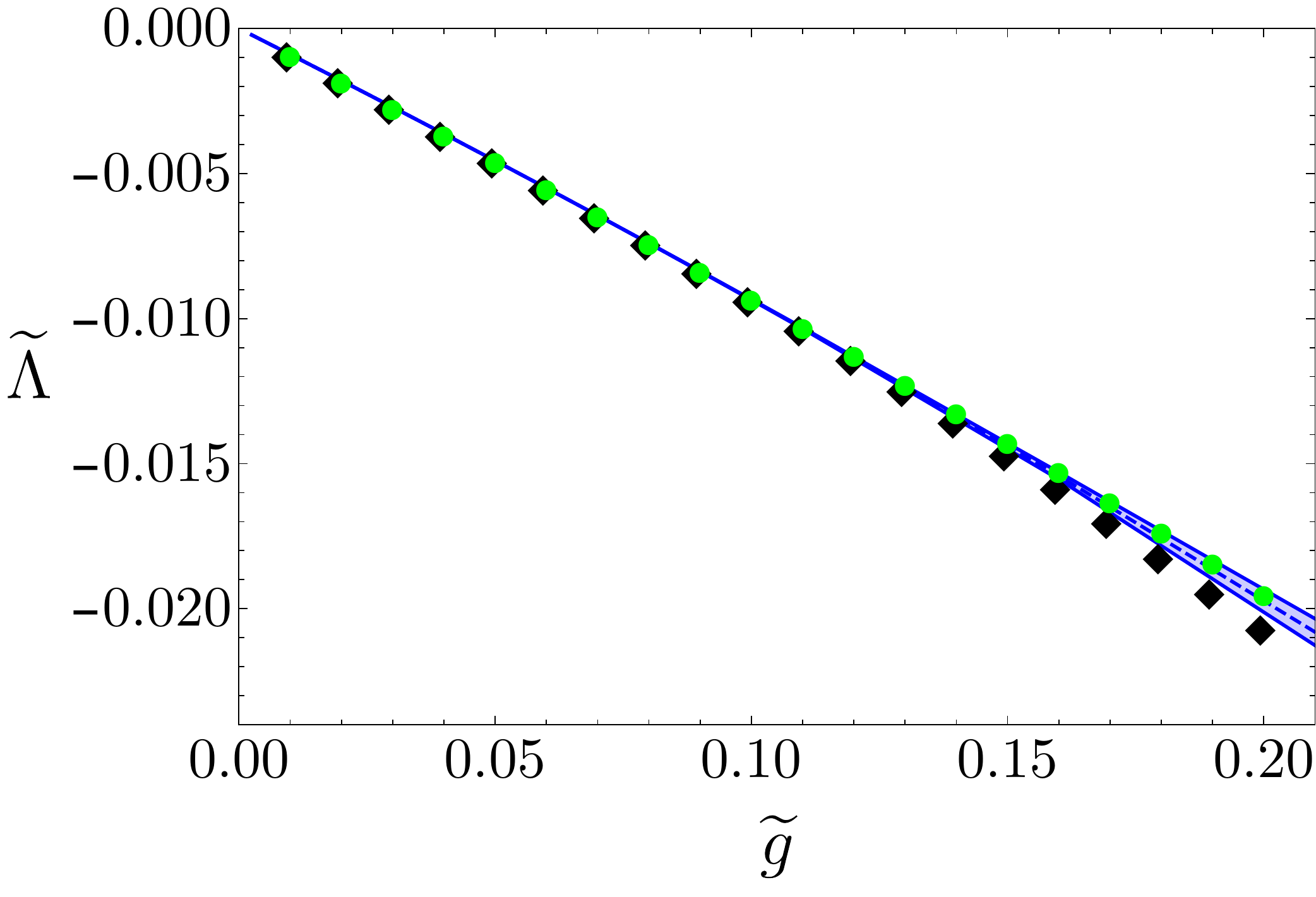}~~~
		\includegraphics[width=.46\textwidth]{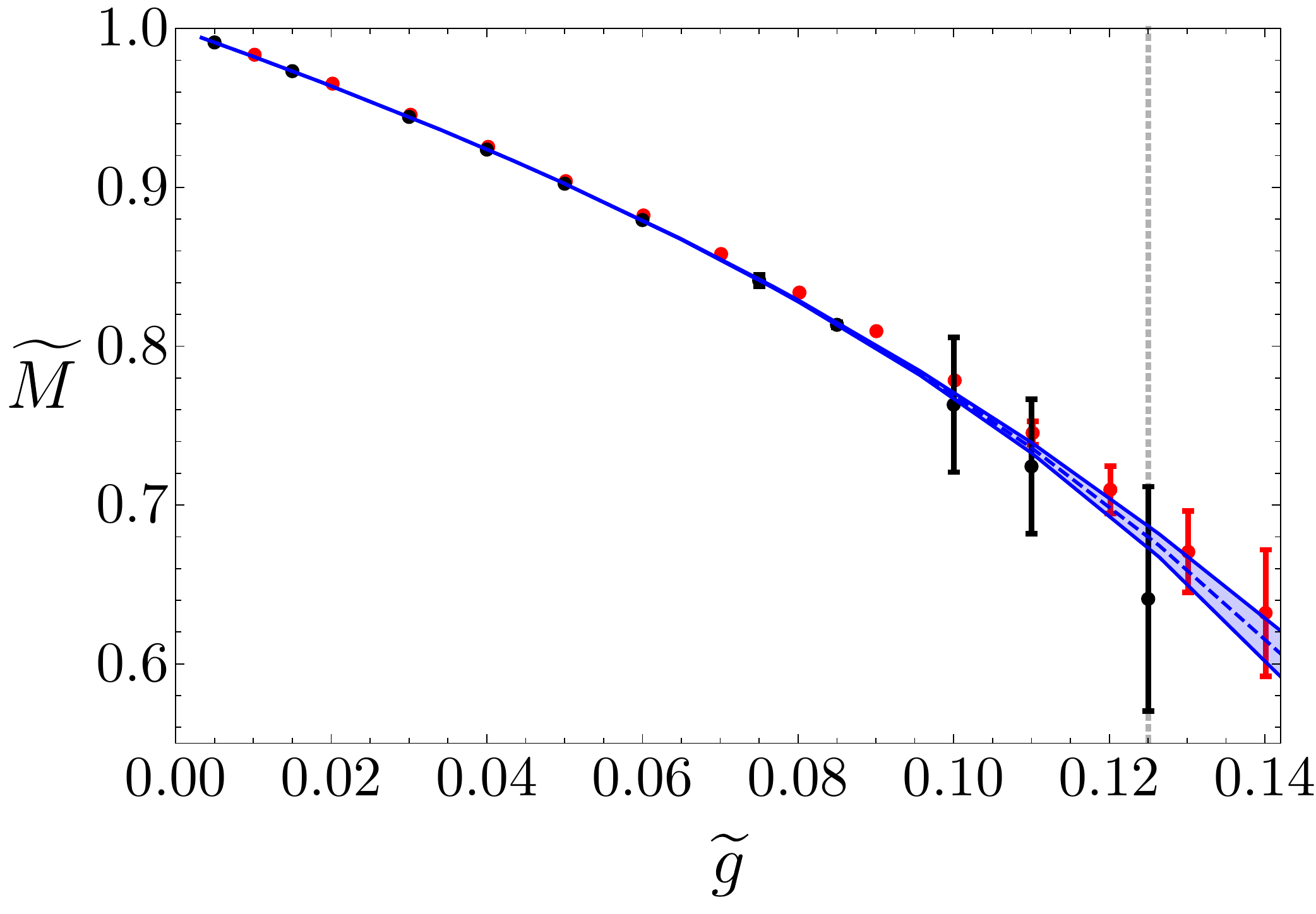}
		\caption{Left panel: comparison with ref.~\cite{Rychkov:2015vap} of the vacuum energy as a function of $\gt$. The blue dashed line are our results obtained by Borel resummation with conformal mapping,
			black and green points are the results of ref.~\cite{Rychkov:2015vap} at finite volume $L=12$  and $L=20$, respectively. Right panel: comparison with refs.~\cite{Rychkov:2015vap,Bajnok:2015bgw} of the
			physical mass as a function of $\gt$. The blue dashed line are our results obtained by Borel resummation with conformal mapping, the black points are the results of ref.~\cite{Bajnok:2015bgw} extrapolated at infinite volume, the red points are the results of ref.~\cite{Rychkov:2015vap} at finite volume $L=20$.
			The vertical dotted line at $\gt = \gt_{k \bar k}\approx 0.125$ is the coupling (taken from ref.~\cite{Bajnok:2015bgw}) where the mass of the elementary particle is  twice the mass of the kink and might become unstable.
		}
		\label{fig:comp}
	\end{figure}

	We compare in fig.~\ref{fig:comp} the values of $\widetilde \Lambda$ and $\Mt$ computed respectively in refs.\cite{Rychkov:2015vap} and \cite{Rychkov:2015vap,Bajnok:2015bgw} with our results.\footnote{We thank the authors of refs.\cite{Rychkov:2015vap,Bajnok:2015bgw} for providing us these data.} Note that the results of ref.~\cite{Rychkov:2015vap} have not been extrapolated at infinite volume: for $\widetilde \Lambda$ we plot their points for two different volumes $L=12$ and $L=20$, while the points for $\Mt$ are at $L=20$.
	In tab.~\ref{tab:comp} we make the comparison explicit for some values of the coupling $\gt$. The values we report for $\Lt$ as computed by ref.~\cite{Rychkov:2015vap} are obtained as the means of the values at two renormalization scales $\mu = 0.9 \, \mt$ and $\mu = 1.1 \, \mt$ and the reported error is the semidifference. The values of $\Mt$ taken from ref.~\cite{Bajnok:2015bgw} have been normalized accordingly to our definitions. As it can be seen, we get more accurate results than those of refs.\cite{Rychkov:2015vap,Bajnok:2015bgw} and they are all in good agreement among themselves.\footnote{The error bars in the data of ref.~\cite{Rychkov:2015vap} for $\Mt$ might not fully take into account truncation effects, explaining the slight disagreement between our results and those of ref.~\cite{Rychkov:2015vap} around $\gt=0.09$.}

	\begin{table}[t!]
		\centering
		\begin{tabular}{ccllll}
			\toprule
			 & $\gt$
			 & \hspace*{20pt}$0.03$               %
			 & \hspace*{20pt}$0.05$               %
			 & \hspace*{20pt}$0.10$               %
			 & \hspace*{20pt}$0.15$
			\\
			\midrule
			 & ref.~\cite{Rychkov:2015vap} $L=12$
			 & $-0.002710(4)$                     %
			 & $-0.004558(6)$                     %
			 & $-0.009340(5)$                     %
			 & $-0.01819(1)$
			\\
			$\Lt$
			 & ref.~\cite{Rychkov:2015vap} $L=20$
			 & $-0.002706(7)$                     %
			 & $-0.00455(1)$                      %
			 & $-0.00929(2)$                      %
			 & $-0.01732(6)$
			\\
			 & This work
			 & $-0.00271009(5)$                   %
			 & $-0.0045602(5)$                    %
			 & $-0.00935(1)$                      %
			 & $-0.0175(2)$
			\\
			\midrule
			\multirow{2}{*}{$\Mt$}
			 & ref.~\cite{Bajnok:2015bgw}
			 & $\phantom{-}0.9444(4)$             %
			 & $\phantom{-}0.9023(5)$             %
			 & $\phantom{-}0.76(4)$               %
			\\
			 & This work
			 & $\phantom{-}0.944401(5)$           %
			 & $\phantom{-}0.90233(5)$            %
			 & $\phantom{-}0.769(2)$              %
			\\
			\bottomrule
		\end{tabular}
		\caption{The values of $\Lt$ and $\Mt$ for some values of $\gt$ and comparison with refs.~\cite{Rychkov:2015vap} and \cite{Bajnok:2015bgw} respectively. The results of ref.~\cite{Rychkov:2015vap} are at finite volume  $L$ and the reported error is obtained as the semidifference between the results at two renormalization points, see the text for additional details. }
		\label{tab:comp}
	\end{table}

	For $\gt_{k\bar k}> \gt$, there might be subtleties in the interpretation of $\Mt$ related to spontaneous symmetry breaking.
	Indeed, the proper way to select a vacuum when the $\mathbb Z_2$  symmetry is spontaneously broken is achieved by adding a small explicit breaking term such as $\epsilon \phi$ and take the limit
$\epsilon \rightarrow 0$ only after $L\rightarrow \infty$. The situation considered in refs.\cite{Rychkov:2015vap,Bajnok:2015bgw} corresponds to the opposite order of  limits, $\epsilon \rightarrow 0$ first and $L\rightarrow \infty$ after, since no breaking term was present to begin with.
	As usual in perturbative QFT, we performed a selection of the vacuum ``by hand", by choosing to expand around any of the two vacua, neglecting the effect of the other.
	There is no need to add an explicit breaking term, so our configuration is equivalent to having taken $L\rightarrow \infty$ first, since we have infinite volume to start with,
	and $\epsilon \rightarrow 0$ after. Since the two limits do not commute, the results for the particle decay found in ref.~\cite{Bajnok:2015bgw} do not obviously apply in our context.
	When $L\rightarrow \infty$, at finite $\epsilon$,  the non-trivial topological Hilbert space sector containing an odd number of kink and anti-kink states decouple.
	While topologically trivial states of kink anti-kink do not decouple, the fundamental particle can no longer kinematically
	decay into freely moving kink and anti-kink states, since single kink and anti-kink states are no longer in the spectrum. In other words, the $\phi$-particle for any finite $\epsilon$ would behave as a spatially extended but stable bound state.
	As $\epsilon$ becomes smaller and smaller, this state becomes less and less bound and for $\epsilon=0$ and $\gt \geq \gt_{k\bar k}$ it unbounds to a pair of essentially free kink-anti-kink states.
	In this case  our results for $\Mt$ do not have a clear interpretation, since they have been obtained assuming the existence
	of a pole of the two-point function. But in the topological trivial sector no single particle state would remain, and the operator $\phi$ would only create multi-particle states.
	In other words, the pole of the two-point function would dissolve in a branch-cut singularity for $\gt \geq \gt_{k\bar k}$.
	In this case $\Mt$ would be related to the threshold energy for the multi particle production, or perhaps it would simply be an unphysical analytic continuation with no obvious significance.
	As we will see in the next subsection, the second hypothesis is favored by our analysis.

	\subsection{Kink States from the Unbroken Phase?}
	\label{subsec:masskink}
	\begin{figure}[t!]
		\centering
		\includegraphics[width=.6\textwidth]{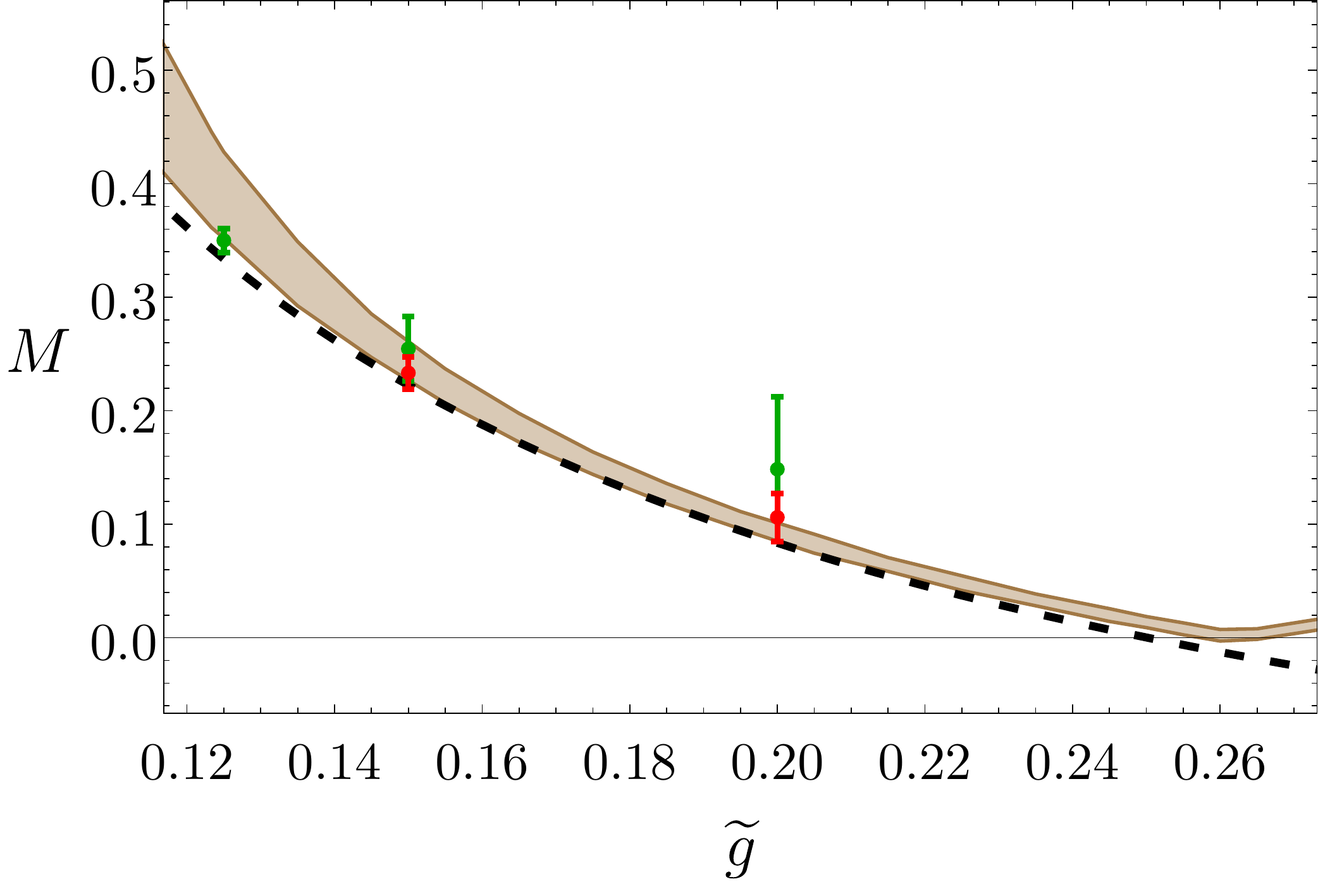}
		\caption{Comparison between the kink mass and the mass $|M|$ in the unbroken phase for $g>g_c$ as a function of $\gt$ in the weakly coupled branch.
			In brown we have $|M|$ as obtained from the Borel resummation of the perturbative series in the unbroken phase up to order $g^8$ using conformal mapping.
			The black dashed line is the kink mass as given by the semiclassical equation (\ref{eq:mkink}). The green and red points are the results obtained by ref.~\cite{Bajnok:2015bgw} from the splitting and antiperiodic sector respectively. See the main text for further details.}
		\label{fig:mkink}
	\end{figure}
	As we mentioned in the last subsection, we cannot have a direct access to single kink and anti-kink states starting from a vacuum where spontaneous symmetry breaking occurs.
	Interestingly enough, however, we might possibly access the kink sector of the theory starting from the unbroken theory with $m^2>0$!
	Indeed, it has been conjectured in ref.~\cite{Serone:2018gjo} that the Borel resummation of correlation functions starting from the unbroken phase reconstructs for $g> g_c$ the correlation functions in a vacuum
	linear combination of $|+\rangle$ and $|-\rangle$ connected by kink configurations, where $\langle \phi\rangle =0$ and cluster decomposition is violated, that is the configuration in refs.\cite{Rychkov:2015vap,Bajnok:2015bgw}.
	A consequence of this conjecture is that the vacuum energy $\Lambda$ for $g>g_c$ should be identified with the vacuum energy $\widetilde \Lambda$ as computed in the weakly and strongly coupled branches.
	More interestingly, $|M(g)|$ as computed in ref.~\cite{Serone:2018gjo} for $g\gtrsim g_c$ should be identified with the mass gap in the non-clustered vacuum, which is given by the kink mass.
	The kink is indeed the lightest single particle excitation close to the phase transition (in fact, the only one when $g$ is sufficiently close to $g_c$).
	The kink mass as a function of $\gt$ has been numerically studied in ref.~\cite{Bajnok:2015bgw}. It turns out that the semi-classical kink mass formula, including one-loop corrections,
	\be
	\frac{M_{{\rm kink}}}{\mt} =\frac{1}{12 \gt} - \frac{3}{2 \pi} + \frac{1}{4 \sqrt{3}}\,,
	\label{eq:mkink}
	\ee
	is in very good agreement with the numerical results for all values of $\tilde g$ in the weakly coupled branch.\footnote{In light of our results this is perhaps not that surprising, since we have shown the validity of perturbation theory in the weakly coupled branch.} In fig.~\ref{fig:mkink} we compare the results of ref.~\cite{Bajnok:2015bgw} with eq.~(\ref{eq:mkink}) and our results for $|M|$ as computed in ref.~\cite{Serone:2018gjo} and analytically continued beyond the phase transition.
	\begin{figure}[t!]
		\centering
		\includegraphics[width=.6\textwidth]{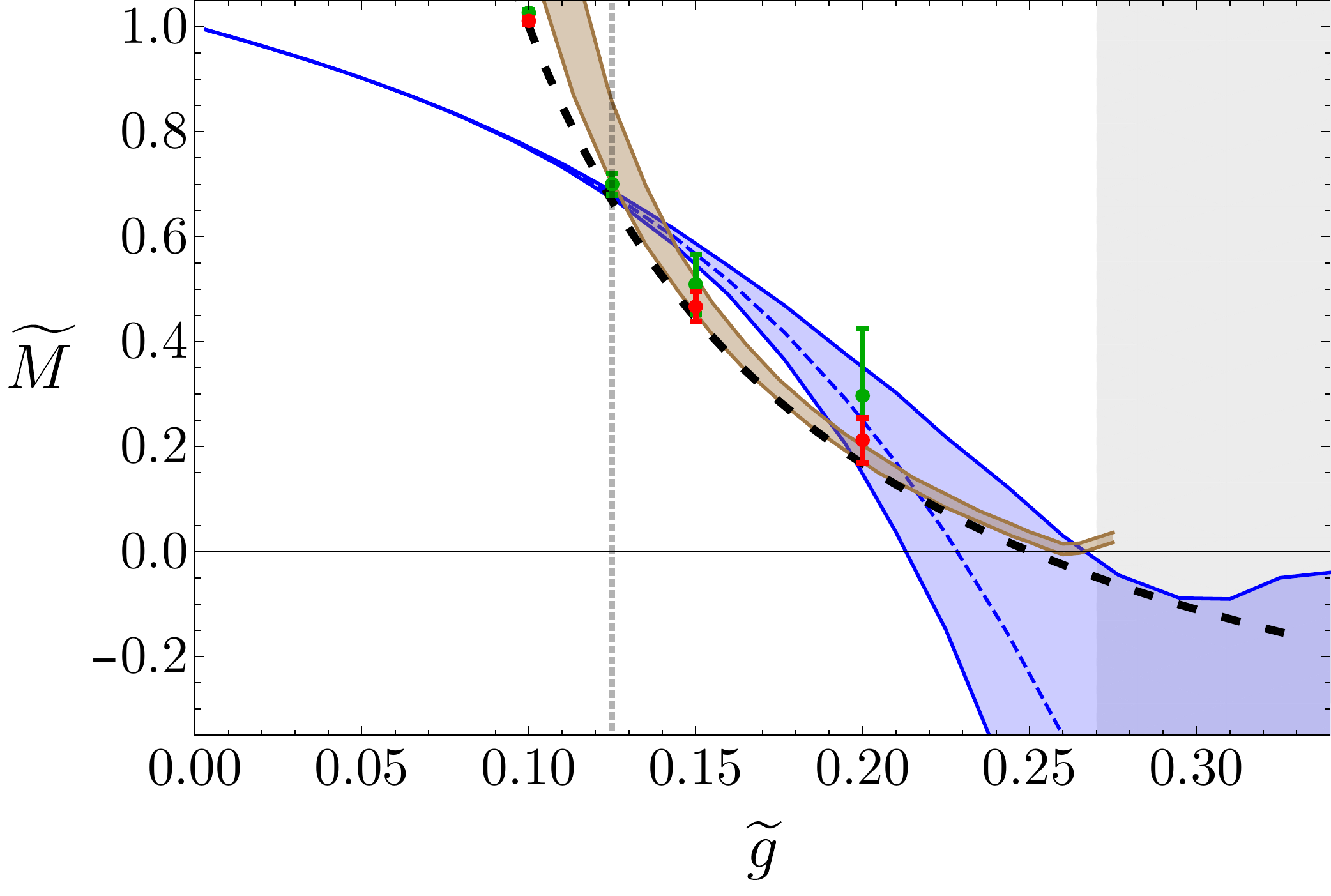}
		\caption{Comparison between $2M_{{\rm kink}}$ and $\widetilde M$ as a function of $\gt$ in the weakly coupled branch.
			The former has been obtained multiplying by a factor 2 the data in fig.~\ref{fig:mkink}, while the latter using conformal mapping in the broken phase, as in fig.~\ref{fig:mass}.
			The vertical dotted line at $\gt = \gt_{k \bar k}\approx 0.125$ is the coupling (taken from ref.~\cite{Bajnok:2015bgw}) where the mass of the elementary particle is  twice the mass of the kink and becomes unstable. The color coding is the same as in figs.~\ref{fig:mass} and \ref{fig:mkink}.}
		\label{fig:Mtilde_kk}
	\end{figure}
	The red and green points are the results of ref.~\cite{Bajnok:2015bgw}. The former are directly obtained by computing the vacuum energy in the anti-periodic sector, while the latter are obtained by computing the energy splitting between the first two states in the topological trivial sector.\footnote{The latter method requires also
		the use of eq.~(\ref{eq:mkink}).} Note that the region of $|M(g)|$ beyond the transition for $g> g_c$ correspond to $\gt< \gt_c^{(w)}$ and that we are using units where $\widetilde m^2=1$,
	while in ref.~\cite{Serone:2018gjo} we had $m^2=1$. The results are in very good agreement with each other, providing evidence to our conjecture.
	The relation between perturbative and non-perturbative states in different phases of a theory is typical in theories enjoying duality symmetries, but
	we are unaware of relations of this sort that involve correlation functions in a vacuum that does not satisfy cluster decomposition.

	We conclude by comparing $\widetilde M$ and $2 M_{{\rm kink}}$ as a function of $\widetilde g$, see fig.~\ref{fig:Mtilde_kk}.
	This is useful because if $\widetilde M$ still describes the mass gap in the theory it should follow the curve $2M_{{\rm kink}}$.
	Though not conclusive, fig.~\ref{fig:Mtilde_kk} does not support this hypothesis and seems  to suggest instead that for $\gt \geq \gt_{k\bar k}$ $\widetilde M$ is an analytic continuation
	with no obvious interpretation.

	\section{Conclusions}
	\label{sec:conclu}

	In this paper we have studied the 2d $\phi^4$ theory in the broken phase, recently shown to be Borel resummable \cite{Serone:2018gjo}.
	We have computed the leading finite action complex instanton solutions, important to determine the large order behavior of the perturbative series
	and to Borel resum it using a generalized conformal mapping method.
	We have computed the perturbative series expansion for the first Schwinger functions  up to order $\gt^4$ and
	Borel resummed the truncated series using our generalized conformal mapping technique.
	The results of the resummation are not as accurate as in the unbroken phase, but allow us to establish that the weakly coupled branch of the broken phase is almost entirely within
	the perturbative regime. This somewhat unexpected result is fully confirmed by comparing our perturbative (resummed or not) results with hamiltonian truncation methods.

	We have also used EPT to compute the vacuum energy at strong coupling, and compared the results
	to the ones obtained in the weakly coupled branch and in the strongly coupled unbroken phase, proving in this way
	Chang duality and the Borel summability of the theory.
	We have finally provided a numerical evidence that the mass gap analytically continued from the unbroken to the broken $\mathbb Z_2$ phase can be identified with a kink state.
	This result is in agreement with the expectation that the analytically continued Schwinger two-point function in the broken phase corresponds to the mass gap in a inhomogeneous vacuum where cluster decomposition does not hold. It would be nice to have a deeper understanding of this phenomenon and to establish if and to what extent it applies for other Schwinger functions and for other Borel resummable theories undergoing phase transitions.

	\section*{Acknowledgments}

	We thank Z. Bajnok, M. Lajer, S. Rychkov and L. Vitale for sharing with us the data
	needed to compare our results with those in refs.~\cite{Rychkov:2015vap,Bajnok:2015bgw}  in figs.~\ref{fig:comp} and ~\ref{fig:mkink}. We thank S.~Cecotti, J.~Elias-Mir\`o, E.~Katz, R.~Konik, M.~Lajer, G.~Mussardo, L.~Vitale and S.~Rychkov for useful discussions. We also thank Z. Bajnok, J.~Elias-Mir\`o and S.~Rychkov for comments on the manuscript.
	G.S.~thanks P.~Lepage for the precious insights in the usage of the VEGAS algorithm.
	Preliminary results of this work have been presented by M.S. at the IHES workshop
	``Hamiltonian methods in strongly coupled Quantum Field Theory", January 8-12, 2018 and at the Caltech workshop``Bootstrap 2018", July 2-27. We thank all the participants of both workshops for useful feedback.
	M.S. acknowledges support from the Simons Collaboration on the Non-perturbative Bootstrap.
	G.S. acknowledges support from the Paris Île-de-France region in the framework of DIM SIRTEQ.

	\appendix

	\section{Series Coefficients of the $0$, $1$ and $2$ Point Functions}
	\label{sec:appendix}

	In this appendix we report the coefficients for the series expansion of the $0$, $1$ and $2$-point function with independent cubic and quartic couplings $\lambda_3$ and $\lambda$. When performing ordinary perturbation theory with $\lambda_3 =\sqrt{2 \lambda}\mt$, the $\lambda^n$ terms with $n>4$ should not be included in the series, since they would require to add terms up to $\lambda_3^{2n}$ that we have not computed.

	In tab.~\ref{tab:Lambda-coeff} we list the coefficients for the vacuum energy up to eight total vertices.
	In tab.~\ref{tab:vev-coeff} we list the coefficients of the VEV of $\phi$ up to eight total vertices.
	In tab.~\ref{tab:Gamma2n-coeff} we list the coefficients $b_{k,l}^{(n)}$ of the $n$th-derivative of the $2$-point function $\widetilde \Gamma _2 ^{(n)}$ at momentum $p^2=-\Mt^2$ defined as
	\begin{equation}
		\widetilde \Gamma_2^{(n)}(-\mt^2) =\mt^{2-2n}  \sum_{k=0,l=0}  b_{k,l}^{(n)}\,\, \left(\frac{\lambda}{\mt^2}\right)^k \left(\frac{\lambda_3}{\mt^2}\right)^l , \quad \quad b_{0,0}^{(n)}=\delta_{n,1}\,.
		\label{eq:Gamma2n-expansion}
	\end{equation}
	Using the procedure explained in ref.~\cite{Serone:2018gjo} and the coefficients $b_{k,l}^{(n)}$  one can determine the perturbative series for the physical mass $\Mt^2$.
\begin{table}[h]
	\centering
	\begin{small}
		\setlength{\tabcolsep}{4pt}
		\begin{tabular}{c|ccccc}
			\toprule
			${\Lt}$     & $\lambda_3^{~0}$          & $\lambda_3^{~2}$                                                     & $\lambda_3^{~4}$   & $\lambda_3^{~6}$  & $\lambda_3^{~8}$ \\
			\midrule
			$\lambda^0$ & 0                         & $-\frac{1}{2}\left(\frac{\psi^{(1)}(1/3)}{4\pi^2}  -\frac 16\right)$ & $-0.037804619(13)$ & $-0.14168986(85)$ & $-0.7158909(36)$ \\
			$\lambda^1$ & 0                         & $0 .079959370431$                                                    & $0 .45728168(71)$  & $3.052200(15)$                       \\
			$\lambda^2$ & $-21 \zeta (3)/16 \pi ^3$ & $-0.37556393(34)$                                                    & $-4.103721(35)$    & $-44.760(19)$                        \\
			$\lambda^3$ & $27 \zeta (3)/8 \pi ^4$   & $1.7780406(30)$                                                      & $34 .6608(25)$                                            \\
			$\lambda^4$ & $-0.116125964(91)$        & $-9.413933(23)$                                                      & $-294.714(45)$                                            \\
			$\lambda^5$ & $0 .3949534(18)$          & $55 .2353(57)$                                                                                                                   \\
			$\lambda^6$ & $-1.629794(22)$           & $-356.38(36)$                                                                                                                    \\
			$\lambda^7$ & $7 .85404(21)$                                                                                                                                               \\
			$\lambda^8$ & $-43.1920(21)$                                                                                                                                               \\
			\bottomrule
		\end{tabular}
	\end{small}
	\caption[]{Perturbative coefficients for the vacuum energy $\Lt$ with independent cubic and quartic coupling $\lambda_3$ and $\lambda$ up to eight total vertices. We set $\mt = 1$ to avoid clutter.\footnotemark}
	\label{tab:Lambda-coeff}
\end{table}
\footnotetext{We thank S.~Rychkov and L.~Vitale for spotting a missing factor 1/2 in the first $\lambda_3^2$ coefficient in a previous version of the paper.}

\begin{table}[h]
	\centering
	\begin{small}
		\setlength{\tabcolsep}{6pt}
		\begin{tabular}{c|ccccc}
			\toprule
			${\langle \phi \rangle}$ & $\lambda_3^{~1}$                           & $\lambda_3^{~3}$   & $\lambda_3^{~5}$ & $\lambda_3^{~7}$ \\
			\midrule
			$\lambda^0$              & 0                                          & $-0.267173395(10)$ & $-1.0631775(38)$ & $-5.773220(83)$  \\
			$\lambda^1$              & $\frac{\psi^{(1)}(1/3)}{\pi^2}  -\frac 23$
			                         & $2 .3297864(28)$                           & $18 .72732(33)$    & $165 .1097(19)$                     \\
			$\lambda^2$              & $-0.94497557(60)$                          & $-17.077459(80)$   & $-235.4280(47)$  &                  \\
			$\lambda^3$              & $3 .795830(20)$                            & $123 .0864(36)$    & $2657 .933(30)$  &                  \\
			$\lambda^4$              & $-17.07032(12)$                            & $-916.534(22)$     &                  &                  \\
			$\lambda^5$              & $87 .5081(29)$                             & $7168.4(6.8)$      &                  &                  \\
			$\lambda^6$              & $-501.799(48)$                             &                    &                  &                  \\
			$\lambda^7$              & $3193 .26(51)$                             &                    &                  &                  \\
			\bottomrule
		\end{tabular}
	\end{small}
	\caption{Perturbative coefficients for the VEV with independent cubic and quartic coupling $\lambda_3$ and $\lambda$ up to eight total vertices. We set $\mt = 1$ to avoid clutter.}
	\label{tab:vev-coeff}
\end{table}
\begin{table}[t]
	\centering
	\begin{small}
		\setlength{\tabcolsep}{6pt}
		\begin{tabular}{c|cccccc}
			\toprule
			${b^{(0)}}$                  & $\lambda_3^{~0}$                               & $\lambda_3^{~2}$                       & $\lambda_3^{~4}$ & $\lambda_3^{~6}$ & $\lambda_3^{~8}$ \\
			\midrule
			$\lambda^0$                  & 0                                              & $-\sqrt{3}$                            & $-6.67024(46)$   & $-40.88772(33)$  & $-326.666(51)$   \\
			$\lambda^1$                  & 0                                              & $10.70608065292$                       & $102 .4789(74)$  & $1123.600(45)$                      \\
			$\lambda^2$                  & $-3/2$                                         & $-61.50066(78)$                        & $-1183.464(99)$  & $-20273(20)$     &                  \\
			$\lambda^3$                  & $\frac{63 \zeta (3)}{2 \pi ^3}+\frac{9}{\pi }$ & $381 .343(23)$                         & $12449.2(1.4)$   &                                     \\
			$\lambda^4$                  & $-14.777287(22)$                               & $-2546.85(11)$                         & $-127782(36)$                                          \\
			$\lambda^5$                  & $66.81651(43)$                                 & $18240.5(2.5)$                                                                                  \\
			$\lambda^6$                  & $-353.2405(28)$                                & $-139410(49)$                                                                                   \\
			$\lambda^7$                  & $2111.715(36)$                                                                                                                                   \\
			$\lambda^8$                  & $-13994.24(54)$                                                                                                                                  \\
			\bottomrule
			\rule{0pt}{11pt} ${b^{(1)}}$ & $\lambda_3^{~0}$                               & $\lambda_3^{~2}$                       & $\lambda_3^{~4}$ & $\lambda_3^{~6}$                    \\
			\midrule
			$\lambda^0$                  & 1                                              & $\frac{3}{\pi }-\frac{1}{\sqrt{3}}$    & $2 .679229(11)$  & $24.8553(15)$                       \\
			$\lambda^1$                  & 0                                              & $-2.4433766(42)$                       & $-45.69101(80)$                                        \\
			$\lambda^2$                  & $0.08094532639$                                & $17.705277(72)$                        & $570.195(40)$                                          \\
			$\lambda^3$                  & $-0.341795194(75)$                             & $-126.0723(24)$                                                                                 \\
			$\lambda^4$                  & $1.8559406(86)$                                & $924 .083(89)$                                                                                  \\
			$\lambda^5$                  & $-10.83118(19)$                                                                                                                                  \\
			$\lambda^6$                  & $68.3310(29)$                                                                                                                                    \\
			\bottomrule
			\rule{0pt}{11pt} ${b^{(2)}}$ & $\lambda_3^{~0}$                               & $\lambda_3^{~2}$                       & $\lambda_3^{~4}$                                       \\
			\midrule
			$\lambda^0$                  & 0                                              & $\frac{3}{\pi }-\frac{2}{\sqrt{3}}$    & $-2.285492(21)$                                        \\
			$\lambda^1$                  & 0                                              & $1.54401986(31)$                                                                                \\
			$\lambda^2$                  & $-0.0128046736$                                & $-12.10262(11)$                                                                                 \\
			$\lambda^3$                  & $0.079771437(20)$                                                                                                                                \\
			$\lambda^4$                  & $-0.5258941(27)$                                                                                                                                 \\
			\bottomrule
			\rule{0pt}{11pt} ${b^{(3)}}$ & $\lambda_3^{~0}$                               & $\lambda_3^{~2}$                                                                                \\
			\midrule
			$\lambda^0$                  & 0                                              & $\frac{9}{\pi }-\frac{14}{3 \sqrt{3}}$                                                          \\
			$\lambda^1$                  & 0                                                                                                                                                \\
			$\lambda^2$                  & $0.00350654051$                                                                                                                                  \\
			\bottomrule
		\end{tabular}
	\end{small}
	\caption{The coefficients $b^{(n)}_{k,l}$ relevant for the determination of the pole mass $\Mt^2$.}
	\label{tab:Gamma2n-coeff}
\end{table}
\FloatBarrier

\end{document}